\documentclass[11pt]{article}

\usepackage[letterpaper,top=2cm,bottom=2cm,left=2cm,right=2cm,marginparwidth=2cm]{geometry}
\usepackage[font=small,margin=2cm]{caption}
\usepackage{authblk}
\usepackage[colorlinks,citecolor=BlueViolet,filecolor=BlueViolet,urlcolor=BlueViolet,bookmarks=false,hypertexnames=true]{hyperref}

\usepackage{graphicx}
\usepackage{algorithm}
\usepackage{algpseudocode}
\usepackage[dvipsnames]{xcolor}
\usepackage{orcidlink}
\usepackage{fancyvrb}
\usepackage{cite}
\usepackage{enumitem}

\usepackage{amsmath}
\usepackage{amssymb}

\usepackage[capitalise]{cleveref}

\usepackage{qcircuit}
\newcommand{\ket}[1]{| #1 \rangle}
\newcommand{\bra}[1]{\langle #1 |}

\newcommand{\bket}[2]{\langle #1|#2\rangle}
\DeclareMathOperator{\tr}{tr}

\DeclareMathSymbol{\mh}{\mathord}{operators}{`\-}
\usepackage[title]{appendix}
\usepackage{subcaption}
\usepackage{multirow}
\usepackage{etoolbox}

\patchcmd{\appendices}{\quad}{: }{}{}
\usepackage[normalem]{ulem}

\usepackage{hyperref}
\usepackage{endnotes}

\definecolor{daxcolor}{rgb}{0.2, 0.1, 1}

\definecolor{jfcolor}{rgb}{0.4, 0.3, 0.8}

\definecolor{stcolor}{rgb}{0.5, 0.4, 0.3}

\definecolor{jycolor}{rgb}{0.913,0.588,0.478}

\usepackage{amsmath, amssymb, amsfonts}
\usepackage{graphicx}
\begin{document}

\newcommand{\dpc}[1]{\textcolor{blue}{ (DP: #1) }}

\title{Solving Fractional Differential Equations on a Quantum Computer: \\
A Variational Approach}

\author[1]{Fong Yew Leong\,\orcidlink{0000-0002-0064-0118}} 
\author[1,2]{Dax Enshan Koh\,\orcidlink{0000-0002-8968-591X}} 
\author[1]{Jian Feng Kong\,\orcidlink{0000-0001-5980-4140}} 
\author[1,3]{Siong Thye Goh\,\orcidlink{0000-0001-7563-0961}} 
\author[1]{Jun Yong Khoo\,\orcidlink{0000-0003-0908-3343}} 
\author[1]{Wei-Bin Ewe\,\orcidlink{0000-0002-4600-0634}} 
\author[1]{Hongying Li\,\orcidlink{0000-0003-4736-295X}} 
\author[1]{Jayne Thompson} 
\author[4,5,6]{Dario Poletti} 

\affil[1]{\small Institute of High Performance Computing (IHPC), Agency for Science, Technology and Research (A*STAR), 1 Fusionopolis Way, \#16-16 Connexis, Singapore 138632, Singapore}
\affil[2]{\small Science, Mathematics and Technology Cluster, Singapore University of Technology and Design, 8 Somapah Road, Singapore 487372, Singapore}
\affil[3]{\small Singapore Management University,
81 Victoria Street,
188065, Singapore}
\affil[4]{\small Science, Mathematics and Technology Cluster and Engineering Product Development Pillar, Singapore University
of Technology and Design, 8 Somapah Road, 487372 Singapore} 
\affil[5]{\small Centre for Quantum Technologies, National University of Singapore 117543, Singapore } 
\affil[6]{\small MajuLab, CNRS-UNS-NUS-NTU International Joint Research Unit, UMI 3654, Singapore}

\date{}

\maketitle

\begin{abstract}
We introduce an efficient variational hybrid quantum-classical algorithm designed for solving Caputo time-fractional partial differential equations. Our method employs an iterable cost function incorporating a linear combination of overlap history states. The proposed algorithm is not only efficient in time complexity, but has lower memory costs compared to classical methods. Our results indicate that solution fidelity is insensitive to the fractional index and that gradient evaluation cost scales economically with the number of time steps. As a proof of concept, we apply our algorithm to solve a range of fractional partial differential equations commonly encountered in engineering applications, such as the sub-diffusion equation, the non-linear Burgers' equation and a coupled diffusive epidemic model. We assess quantum hardware performance under realistic noise conditions, further validating the practical utility of our algorithm.

\end{abstract}

\section{Introduction}
\label{sec:introduction}

Differential equations play an important role in the modeling of many practical engineering problems. Fractional differential equations, which are generalizations of differential equations to arbitrary non-integer order, involve fractional derivatives such as $\frac{d^\alpha}{dx^\alpha}$, where $\alpha$ is a non-integer real number. These models have been shown to be capable of modeling more complex processes that exhibit memory effects or non-local behavior. For example, fractional differential equations have found applications in fractional advection-dispersion equations~\cite{Liu2017} and viscoelastic flow problems~\cite{Wang2023}. Fractional-order derivatives in time have also been used to model cancer tumor growth, by numerically fitting the fractional order \cite{iyiola2014fractional}.

Here, we consider numerical solutions to time-fractional diffusive differential equations of order $\alpha$, which derive from non-Markovian continuous-time random walks \cite{Lynch2003}. Finite difference methods are effective and simple to implement for time-fractional problems \cite{Li2013}, where implicit methods are preferred to explicit methods for numerical stability \cite{Murio2008}. However, the time-fractional derivative creates a \emph{global dependence} problem \cite{Lin2007}, where previous solutions are accessed from memory, rendering storage expensive for spatially large problems.

Quantum computers can offer new tools for efficiently solving differential equations. In particular, variational quantum algorithms (VQAs) have emerged as a leading strategy in the noisy intermediate-scale quantum (NISQ) era \cite{Bharti2022,cerezo2021variational}, featuring low-depth parameterized quantum circuits, alongside classical optimizers used to minimize a cost function. These VQAs  yield quantum states that encode approximate solutions to partial differential equations. They have been successfully applied to both linear \cite{li2023variational,sato2021variational,ali2023performance} and nonlinear differential equations~\cite{Lubasch2020,sarma2023quantum} in a diverse range of fields, including electromagnetics~\cite{ewe2022variational}, collodial transport~\cite{leong2023variational}, fluid dynamics~\cite{jaksch2023variational,rigas2022variational,Leong2022}, and finance \cite{kubo2022pricing}.

In this paper, we propose and implement a variational quantum optimization scheme for solving time-fractional diffusive partial differential equations in space and time, either implicitly or semi-implicitly, using simple quantum overlap measurements for the time integral term. We show that our approach is not only efficient in time complexity, but can also yield memory cost advantages in handling the global dependence problem. The rest of the paper is organized as follows. In \cref{sec:FDE}, we introduce the time-fractional diffusion equation and present our variational quantum optimization scheme for numerical integration. In \cref{sec:implementation},  we discuss the details of the implementation and evaluate time and memory complexities of our scheme. In \cref{sec:numerial}, we present numerical experiments, including time-fractional Burgers' equation for hydrodynamics (\cref{subsec:burgers}), and fractional diffusive epidemic model (\cref{subsec:epidemic}). Lastly, we trial our algorithm on noise model simulators and hardware in \cref{sec:hardware}, and conclude with discussions in \cref{sec:discussion}.

\section{Fractional diffusion equations}\label{sec:FDE}

We consider the fractional diffusion equation \cite{mainardi2007some, Murio2008} in space $x$ and time $t$,
\begin{align}
    D_t^\alpha u(t,x) = \partial_{xx} u(t,x),
    \label{eq:1}
\end{align}
where $D_t^\alpha$ is the Caputo fractional derivative of the $\alpha$-th order with respect to $t$, defined as
\begin{align}
    \frac{D^\alpha u(t,x)}{\partial {t^\alpha}} = \frac{1}{\Gamma(n-\alpha)}\int^t_0 \frac{\partial^n {u(s,x)}}{\partial s^n} \frac{ds}{(t-s)^{\alpha-n+1}}, \quad n-1<\alpha<n,
\end{align}
where $\Gamma$ denotes the Gamma function and $n = \lceil \alpha \rceil \in \mathbb N$ is the ceiling of $\alpha$.

\subsection{Difference scheme}
We set up a space-time rectangular domain with spatial extent $L$ and temporal duration $T$, dividing it into a regularly spaced grid with integer size $N \times M$. The spatial grid points are defined as $x_i = ih$ where $h=L/N$ and $i = 1,2, \dots, N$, and the temporal grid points as $t_k = k\tau$ where $\tau = T/M$ and $k = 1,2, \dots, M$.

The first-order finite difference of the Caputo derivative of order $(0<\alpha\leq1)$ is approximated by \cite{Li2013,TahaAbdulazeez2022}
\begin{align}
    D_t^\alpha u(t_k,x_n) \simeq g_{\alpha,\tau} \sum_{j=1}^k w_j^{(\alpha)}(u_n^{k-j+1}-u_n^{k-j}),
    \label{eq:caputo_derivative_first_order}
\end{align}
where $u_n^k \equiv u(t_k,x_n)$, $g_{\alpha,\tau} := \tau^{-\alpha}/\Gamma(2-\alpha)$ and $w_j^{(\alpha)} := j^{1-\alpha} - (j-1)^{1-\alpha}$. Rearranging \cref{eq:caputo_derivative_first_order} yields a discrete sum of terms $u_n^j$ as
\begin{align}
    D_t^\alpha u(t_k,x_n) \simeq g_{\alpha,\tau} \left[ u_n^k - w_k u_n^0 + \sum_{j=1}^{k-1} \left(w_{j+1}-w_{j}\right)u_n^{k-j} \right],
    \label{eq:caputo_FD_derivative}
\end{align}
where $w_1 = 1$ by definition.

The finite difference of the spatial derivative is approximated by
\begin{align}
    \partial_{xx} u(t_k,x_n) \simeq \frac{u_{n+1}^k - 2u_{n}^k + u_{n-1}^k}{h^2}.
    \label{eq:spatial_finite_difference}
\end{align}

Together, the difference scheme for the fractional diffusion equation can be iterated in time as
\begin{align}
    \left( 1 - a \delta \right) u_{n}^k
    = w_k u_n^0 - \sum_{j=1}^{k-1} \left(w_{j+1}-w_{j}\right)u_n^{k-j},
\end{align}
where $a = g_{\alpha,\tau}^{-1} h^{-2}$ and $\delta u_n := u_{n+1} -2u_{n} + u_{n-1}$. In a more compact matrix form, this reads
\begin{align}
\label{eq:compact_matrix_form_A}
    A\mathbf{u}^k = w_k \mathbf{u}^0 - \sum_{j=1}^{k-1} \Delta w_{j} \mathbf{u}^{k-j},
\end{align}
where $\mathbf{u}^k := u(t_k,x)$, $\Delta w_{j} := w_{j+1}-w_{j}$, and $A$ is an $N \times N$ symmetric tridiagonal matrix with $b = 1+2a$ on the main diagonal, $-a$ on the adjacent off-diagonals, and terms in the corners of the matrix that are specified by the boundary conditions. More precisely, the matrix $A$ is given by
\begin{align}
A = \begin{bmatrix}
c & -a & 0 & & & \cdots &  & d \\
-a & b & -a & 0 & & \cdots  & & 0 \\
0 & -a & b & -a & & \cdots  & & 0 \\
\vdots &  &  & \ddots  & & & & \vdots \\
0 & & \cdots & & 0 & -a & b & -a  \\
d & & \cdots & & & 0 & -a & c
\end{bmatrix}\in \mathbb{R}^{N \times N} ,
\label{eq:matrix_A}
\end{align}
where $(c,d) = (b,-a)$ for periodic boundary condition,  $(c,d) = (b,0)$ for Dirichlet boundary condition and $(c,d) = (1+a,0)$ for Neumann boundary condition. The right-hand side of \cref{eq:compact_matrix_form_A} carries the history of solutions up to $k-1$. 

\subsection{Variational quantum optimization}
Using Dirac notation, we rewrite the iterative fractional diffusion equation as
\begin{align}
    A \ket{\tilde{u}^k} = \ket{\tilde{f}^{k-1}},
    \label{eq:fde}
\end{align}
where we approximate the unnormalized quantum state $\ket{\tilde{u}^k}$ at time $k$ by an ansatz or trial state formed by a set of parameterized unitaries $R$ and unparameterized entangling unitaries $W$ as,
\begin{align}
    \ket{\tilde{u}^k} := r^k\ket{u^k} = r^k\prod_{i=1}^{l} W_{l-i+1}R(\boldsymbol{\theta}_{l-i+1}^k) \ket{0},
    \label{eq:ansatz}
\end{align}
where $r^k$ is the norm of the unnormalized state $\ket{\tilde{u}^k}$, $\boldsymbol{\theta}$ is a set of gate parameters and $l$ is the number of layers in the ansatz. The unnormalized source state $\ket{\tilde{f}^{k-1}}$ is a linear combination of history states in $[0,k-1]$. 

Following \cite{sato2021variational}, we define a cost function based on potential energy at time $k$,
\begin{align}
    \mathcal{C}^k (r^k,\boldsymbol{\theta}^k) = \frac{1}{2} (r^k)^2 \bra{u^k}A\ket{u^k} - \frac{1}{2} r^k \left( \bket{u^k}{\tilde{f}^{k-1}} + \bket{\tilde{f}^{k-1}}{u^k} \right),
\end{align}
where
\begin{align}
    \ket{\tilde{f}^{k-1}} = w_k r^0\ket{u^0} - \sum_{j=1}^{k-1} \Delta w_{j}  r^{k-j}\ket{u^{k-j}}.
    \label{eq:k_overlap}
\end{align}
With that, $\bket{u^k}{\tilde{f}^{k-1}}$ expands into a list of $k$ measurable overlap terms $\bket{u^k}{u^{[0,k-1]}}$ \cite{comment1}.

Introducing $\ket{f,u} := 2^{-1/2}(\ket{0}\ket{f}+\ket{1}\ket{u})$, we have
\begin{align}
    \mathcal{C}^k (r^k,\boldsymbol{\theta}^k) = \frac{1}{2} (r^k)^2 \bra{u^k}A\ket{u^k} - r^k \left( w_k r^0 \langle u^k,u^0 \rangle
    - \sum_{j=1}^{k-1} \Delta w_{j} r^{k-j} \langle u^k,u^{k-j} \rangle \right).
\end{align}
Here, $\langle f,u \rangle := \bra{f,u}X\otimes \mathbb{I}^{\otimes n}\ket{f,u}$ is used as a notation shorthand, where $X = \ket{1}\bra{0} + \ket{0}\bra{1}$ is the Pauli-X matrix and $\mathbb{I} := \mathbb{I}_0 + \mathbb{I}_1 = \ket{0}\bra{0} + \ket{1}\bra{1}$ is the $2\times 2$ identity matrix. Also, $n = \log_2 N$ denotes the number of qubits representing $N$ spatial grid points.

\begin{figure}
\centering
\includegraphics[width=1.0\linewidth]{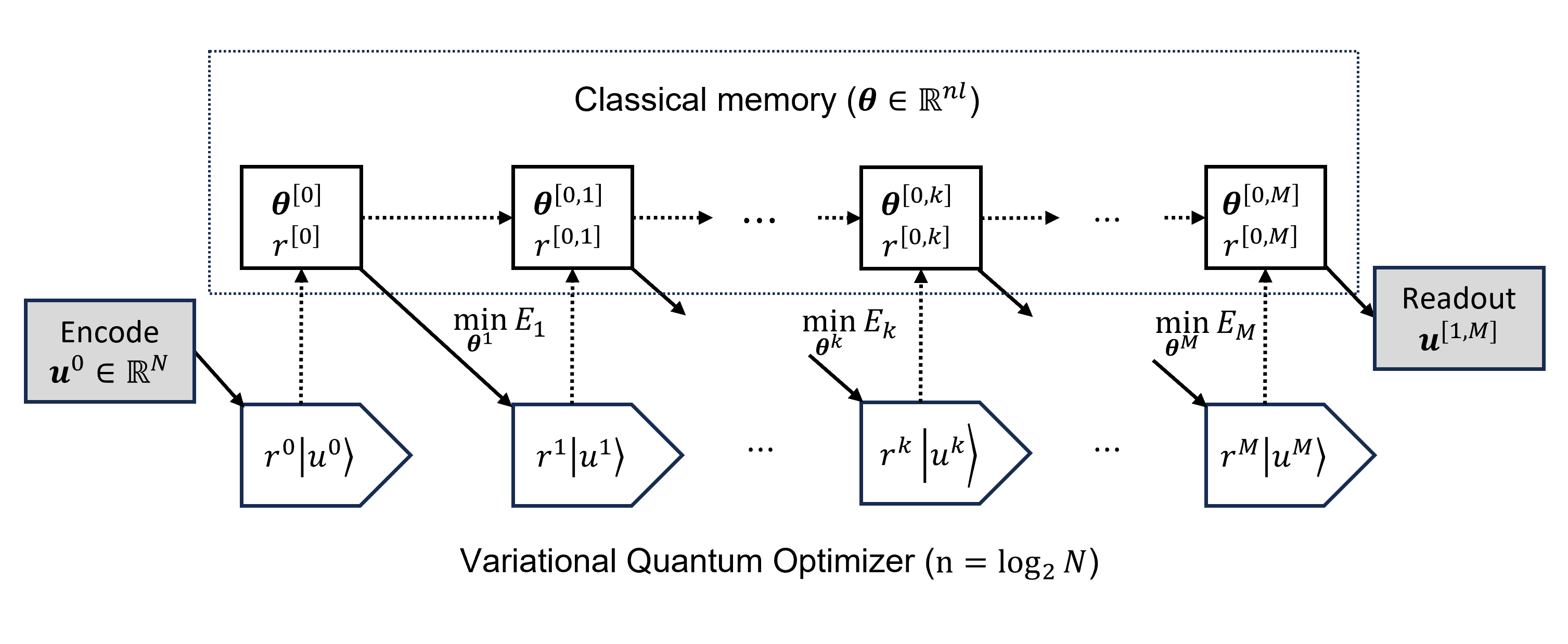}
\caption{Schematic diagram of variational quantum algorithm for time-fractional differential equations. The initial solution, encoded by the vector $\boldsymbol{\theta}^0$ containing $nl$ optimized parameters and the norm $r^0$, is stored in classical memory. Subsequent iterations add $\boldsymbol{\theta}^k$ and $r^k$ to the growing pool of classically stored parameters $\boldsymbol{\theta}^{[0,k-1]}$ and norms $r^{[0,k-1]}$, until the final time-step $M$ is reached. Solutions can be read out selectively via tomography using up to $\mathcal O(n2^n)$ measurements on specific $\boldsymbol{\theta}^k$ parameterized ansatz corresponding to time $k$. Sampling complexity can be reduced using partial tomography or shadow tomography techniques.}
\label{fig:fig0}
\end{figure}

Taking the derivative of $\mathcal{C}^k$ with respect to $r^k$, the optimal norm
\begin{align}
    r^k = \frac{ \langle u^k, \tilde{f}^{k-1} \rangle }
    {\bra{u^k}A\ket{u^k}}
    \label{eq:norm}
\end{align}
can be eliminated from the cost function
\begin{align}
    \mathcal{C}^k (r^k,\boldsymbol{\theta}^k) = 
    -\frac{1}{2} \frac{\left( { \langle u^k, \tilde{f}^{k-1} \rangle } \right)^2} 
    {\bra{u^k}A\ket{u^k}},
    \label{eq:cost_function}
\end{align}
which can be measured using quantum circuits (\cref{fig:figA2}a,b) and optimized classically as
\begin{align}
    \underset{\boldsymbol{\theta}^k}{\arg\min}\  \mathcal{C}^k \left( r^k, \boldsymbol{\theta}^k \right),
    \label{eq:min_cost_function}
\end{align}
via either gradient-based (evaluate $\partial_{\boldsymbol{\theta}^k} \mathcal{C}^k$) or gradient-free methods. A schematic of the proposed variational quantum algorithm is shown in \cref{fig:fig0}.

\section{Implementation}\label{sec:implementation}

\subsection{Hamiltonian decomposition}
The Hamiltonian matrix $A$ (\cref{eq:matrix_A}) can be decomposed into
\begin{align}
A = b\mathbb{I}^{\otimes n} - a \left\{ \underbrace{\mathbb{I}^{\otimes n-1} \otimes X}_{H_{1}} + \mathcal{S}^{\dagger}\big[\underbrace{\mathbb{I}^{\otimes n-1} \otimes X}_{H_{2}} - \underbrace{\mathbb{I}_{0}^{\otimes n-1} \otimes X}_{H_{3}} + \underbrace{\mathbb{I}_{0}^{\otimes n-1} \otimes \mathbb{I}}_{H_{4}}\big]\mathcal{S} \right\}.
\label{eq:A_decomposition}
\end{align}
Since expectation values of the identity operator $\mathbb{I}$ are equal to 1, i.e.\ $\bra \phi \mathbb{I}^{\otimes n} \ket \phi = \bra{\phi'} \mathbb{I}^{\otimes n} \ket{\phi'} = 1$ \cite{sato2021variational}, evaluating the expectation value of $A$ requires only the evaluation of expectation values of the simple terms ($H_{1-2}$ for periodic boundary condition, $H_{1-3}$ Dirichlet boundary condition and $H_{1-4}$ for Neumann boundary condition). The operator $\mathcal{S}$ denotes the \textit{n}-qubit cyclic shift operator,
\begin{align}
    \mathcal{S} = \sum^{2^n-1}_{i=0}\ket{(i+1)\ \text{mod}\ 2^n}\bra i,
    \label{eq:shift_op}
\end{align}
which can be implemented using relative-phase Toffoli gates, Toffoli gates, a \textsc{cnot} gate and an X gate \cite{Maslov2016}. 

\subsection{Ansatz initialization} 

For problems that admit only real solutions, it is preferable to use a real-amplitude ansatz, represented by \cref{eq:ansatz} formed by $l$ repeating blocks, each consisting of a parameterized $R$ layer with one $R_Y(\theta)$ gate on each qubit, followed by unparameterized $W$ layer with a \textsc{cnot} gate between consecutive qubits \cite{alghassi2022}:
\begin{align}
    \ket{u^k} = \prod_{i=l}^1 \left[\prod_{a=n-1}^1 C_a X_{a+1} \bigotimes_{j=1}^n R_Y(\theta_{i,j}) \right]\ket 0,
    \label{eq:HEAnsatz}
\end{align}
where we have employed the product notation with an initial index greater than the stopping index to signify that the terms in the product are arranged in decreasing index order \cite{comment2}. 
In addition, we consider a real-amplitude ansatz with circular entanglement, such that
\begin{align}
    \ket{u^k} = \prod_{i=l}^1 \left[C_nX_0 \prod_{a=n-1}^1 C_a X_{a+1} \bigotimes_{j=1}^n R_Y(\theta_{i,j}) \right]\ket 0, \quad n > 2,
    \label{eq:circular_ansatz}
\end{align}
which has been shown to be efficient for solving partial differential equations \cite{leong2023variational}. See \cref{fig:figA1} for circuit diagrams of the ansaetze.

To solve \cref{eq:fde}, we prepare the initial ansatz state $\ket{\tilde{u}^0} = r^0 \ket{u(\boldsymbol{\theta}^0)}$ using classical optimization on parameters $\boldsymbol{\theta} = \left(\theta_{i,j}\right)_{(i,j) \in \{1,\ldots,l\}\times\{1,\ldots,n\}}$ such that 
\begin{align}
    \boldsymbol{\theta}^0 \in \underset{\boldsymbol{\theta}^0}{\arg\min} \left\| \bket{\hat{u}^0}{u(\boldsymbol{\theta}^0)}
    \right\|,
    \label{eq:encoding}
\end{align}
where $\ket{\hat{u}^0} \equiv \mathbf{u}^0 / \left\|\mathbf{u}^0\right\|$ is the normalized initial classical solution vector and $r^0 = \left\|\mathbf{u}^0\right\|$ is the initial norm.

\subsection{Gradient estimation}
The gradient of the cost function \cref{eq:cost_function} can be estimated on quantum computers using the parameter shift rule. Following \cite{sato2021variational}, here we write down the partial derivative of the cost function with respect to parameter $\theta^k_{i,j}$ indexed by $(i,j) \in [1,nl]$ at $k$, as
\begin{align}
\begin{split}
    \frac{\partial \mathcal{C}^k}{\partial {\theta}_i^k} = 
    - \frac{1}{2} \left( \frac{ \langle u^k, \tilde{f}^{k-1} \rangle } {\bra{u^k}A\ket{u^k}} \right) \left[ w_k r^0 \left\langle \frac{\partial u^k}{\partial \theta_i^k},u^0 \right\rangle  - \sum_{j=1}^{k-1} \Delta w_{j} r^{k-j} \left\langle \frac{\partial u^k}{\partial \theta_i^k},u^{k-j} \right\rangle \right] \\
    + \frac{1}{2} \left( \frac{ \langle u^k, \tilde{f}^{k-1} \rangle } {\bra{u^k}A\ket{u^k}} \right)^2  \left\langle \frac{\partial u^k}{\partial \theta_i^k}, u^k \right\rangle_A,
    \label{eq:cost_derivative}
\end{split}
\end{align}
where $\langle f,u \rangle_A := \bra{f,u}X\otimes A\ket{f,u}$ is used as a notational shorthand. The parametric shift is implemented by a $\pi$ rotation on the $i$-th $R_Y$ gate as
\begin{align}
\frac{\partial u^k}{\partial \theta_i^k} = u^k \left(\theta^{\lfloor i/n \rfloor}_{i\%n} + \pi \right),
\end{align}
where $i\%n$ is the remainder $r\in \{0,1,\ldots, n-1\}$ obtained when $i$ is divided by $n$ and $\lfloor i/n \rfloor$ is the integer part of $i/n$. Thus, for each index $i$ at time $k$, gradient estimation requires at least $k$ overlap measurements.

\subsection{Time complexity and memory scaling}
\emph{Time complexity}. --- Here we briefly estimate the time complexity of the algorithm based on the number of time steps, quantum circuits, gates and shots required, neglecting classical overheads such as parameter update, iteration, functional evaluation and initial encoding. 

For a number of time steps $M$, the overall time complexity is 
\begin{align}
    \mathcal{T} \approx M N_{\mathrm{it}} N_{g}
    \left[ \frac{Ml + (l+n^2)}{\epsilon^2} \right],
\end{align}
where $n$ is the number of qubits, $l$ is the number of ansatz layers, $N_{\mathrm{it}}$ is the mean number of iterations required per time step and $N_{g}$ is the mean number of evaluations required for gradient estimation per iteration. Specifically, the terms in square brackets are the gate complexities for the fractional derivative $Ml$ based on $M/2$ overlap circuits containing $2l$ parametric gates in depth and for the Hamiltonian $(l+n^2)$ based on $\mathcal{O}(1)$ circuits ($2-4$ \cref{eq:A_decomposition}) containing $l$ parametric gates in depth for the ansatz and $\mathcal{O}(n^2)$ non-parametric gates for the shift operator. Each circuit is sampled repeatedly for $\mathcal{O}(\epsilon^{-2})$ shots per measurement, where $\epsilon$ is the desired precision. We remark that with quantum amplitude estimation \cite{brassard2002quantum}, the number of shots required may be quadratically reduced from $\mathcal O(\epsilon^{-2})$ to $\mathcal O(\epsilon^{-1})$. This decrease, however, entails increased circuit depth, as exemplified by the $\Omega(\epsilon^{-1})$ circuit depths utilized in \cite{brassard2002quantum,Suzuki_2020,aaronson2020quantum}. Given these challenges in reducing circuit depth in quantum amplitude estimation and its extensions~\cite{tiron2022low,10.1063/5.0042433, wang2021minimizing,johnson2022reducing}, we do not consider this approach in the present study, deferring it for future exploration.

For $n \sim l$, the gradient-based optimizer $N_g \sim \mathcal{O}(nl)$ yields an overall time complexity 
\begin{align}
    \mathcal{T} \sim \mathcal{O} \left( \max \left\{ \frac{N_{\mathrm{it}} M^2nl^2}{\epsilon^2},\frac{N_{\mathrm{it}} Mn^3l}{\epsilon^2} \right\} \right),
    \label{eq:time_scaling}
\end{align}
scaling as $\mathcal{O} (M^2)$ for overlap circuits, or $\mathcal{O} (M)$ for shift operators, the latter scaling applicable to the non-fractional diffusion/heat equation ($\alpha = 1$, \cite{sato2021variational}). Using simultaneous perturbation stochastic approximation (SPSA) \cite{spall1992multivariate}, only two circuit executions are required per gradient evaluation $N_g \sim \mathcal{O}(1)$, independent of the number of parameters, similar to gradient-free optimization.

The present algorithm is efficient in time complexity with respect to number of spatial grid points $N(=2^n)$ \cite{sato2021variational,Sato2024}, neglecting initial encoding costs. Quantum advantage with respect to time $T$ may be achievable in higher dimensions (Remark 2 \cite{Sato2024}, Remark 3 \cite{Hu2024}).

\emph{Memory scaling}. --- To solve the Caputo derivative \cref{eq:caputo_FD_derivative}, a classical algorithm assesses $\mathcal{O}(kN)$ parameters from memory at time $k$, compared to $\mathcal{O}(knl)$ parameters for variational quantum algorithm \cref{eq:k_overlap}. Assuming $n \sim l$, the memory cost of the Caputo sums scales sub-exponentially as $\mathcal{O}(k\log N)$ (\cref{fig:fig0}), demonstrating efficient algorithmic space complexity.

\section{Numerical experiments}\label{sec:numerial}

The circuits are implemented in the quantum software framework \emph{Pennylane} \cite{bergholm2018pennylane} using noiseless statevector emulation on \textsc{lightning} qubit backend. Parametric updates are performed using the limited-memory Broyden-Fletcher-Goldfarb-Shanno boxed (L-BFGS-B) optimizer with relative tolerance of $10^{-6}$ and gradient tolerance of $10^{-6}$. The L-BFGS-B optimizer evaluates gradients using finite difference without explicit gradient inputs from \cref{eq:cost_derivative}. Initial encoding \cref{eq:encoding} is performed using SciPy's \textsc{minimize} function with an initial null parameter set as $(0 ,\dots, 0)$.

\subsection{Sub-diffusion equation} \label{subsec:subdiffusion}
We solve for time-fractional $(0 < \alpha \leq 1)$ sub-diffusion equation \cref{eq:1} as an initial value problem with initial condition $u(0,x) = x(1-x)$ and Dirichlet boundary condition $u(t,0) = u(t,1) = 0$ \cite{Murio2008}. 

\begin{figure}
\centering
\includegraphics[width=1.0\linewidth]{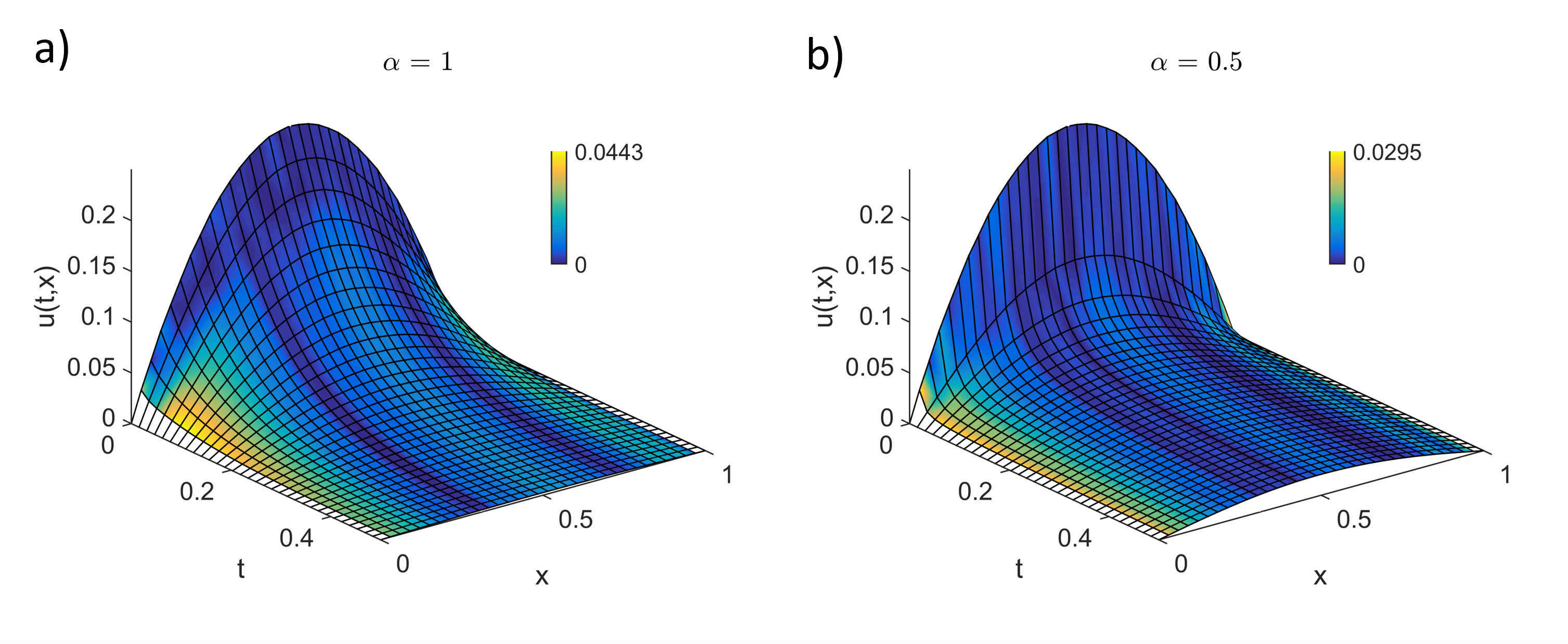}
\caption{Variational quantum solutions to time-fractional diffusion equation in $N \times M = 32 \times 32$ space-time domain with qubit-layer count $(n,l)=(5,4)$, for (a) $\alpha = 1$ and (b) $\alpha = 0.5$. Initial and boundary conditions are $u(0,x) = x(1-x)$ and $u(t,0) = u(t,1) = 0$. Colorbar denotes relative deviation from classical finite difference solution on the same domain.
} 
\label{fig:fig1}
\end{figure}

\begin{figure}
    \centering
    \includegraphics[width=1.0\linewidth]{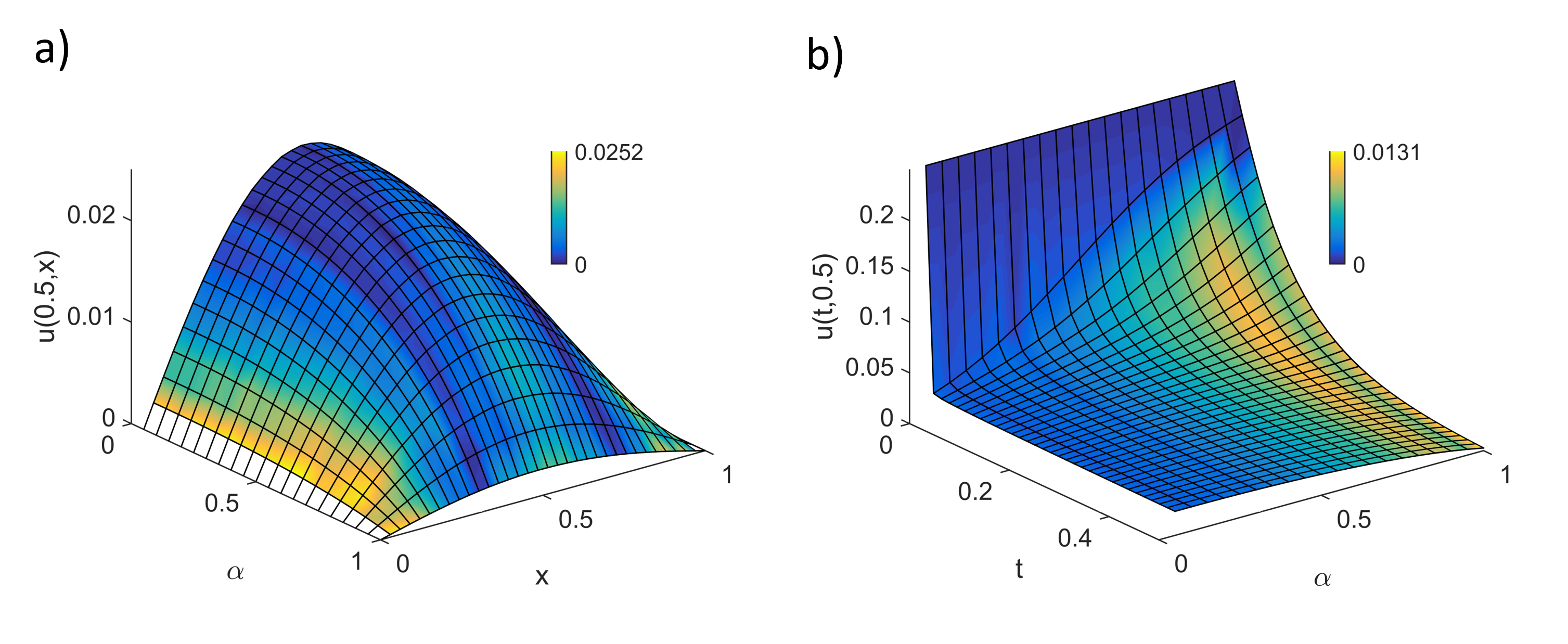}
    \caption{Variational quantum solutions (as Fig.~\ref{fig:fig1}) plotted on (a) $\alpha - x$ at $t = 0.5$, and (b) $t - \alpha$ at $x = 0.5$. Colorbar denotes relative deviation from classical finite difference solution on the same domain.}
    \label{fig:fig2}
\end{figure}

Figure \ref{fig:fig1} shows that the variational quantum solutions agree with classical sub-diffusion solutions in $32 \times 32$ space-time domain with qubit-layer count $(n,l)=(5,4)$, for (a) $\alpha = 1$ and (b) $\alpha = 0.5$. Figure~\ref{fig:fig2} shows that low fractional-order solutions tend to diffuse faster for short times, but slow down for long times, both characteristics of sub-diffusive behavior. 

The trace error at time $k$
\begin{align}
    \epsilon_{\mathrm{tr}}^k = \sqrt{1 - \left\| \bket{\hat{u}^k}{u(\boldsymbol{\theta}^k)}\right\|^2},
    \label{eq:trace_error}
\end{align}
can be averaged over the number of time steps $M$ for $0 < \alpha \leq 1$, where $\ket{\hat{u}^k} \equiv \mathbf{u}^k / \left\|\mathbf{u}^k\right\|$ is the normalized classical solution vector at time $k$.

\subsubsection{Trace error and optimization}
\begin{figure}
    \centering
    \includegraphics[width=1.0\linewidth]{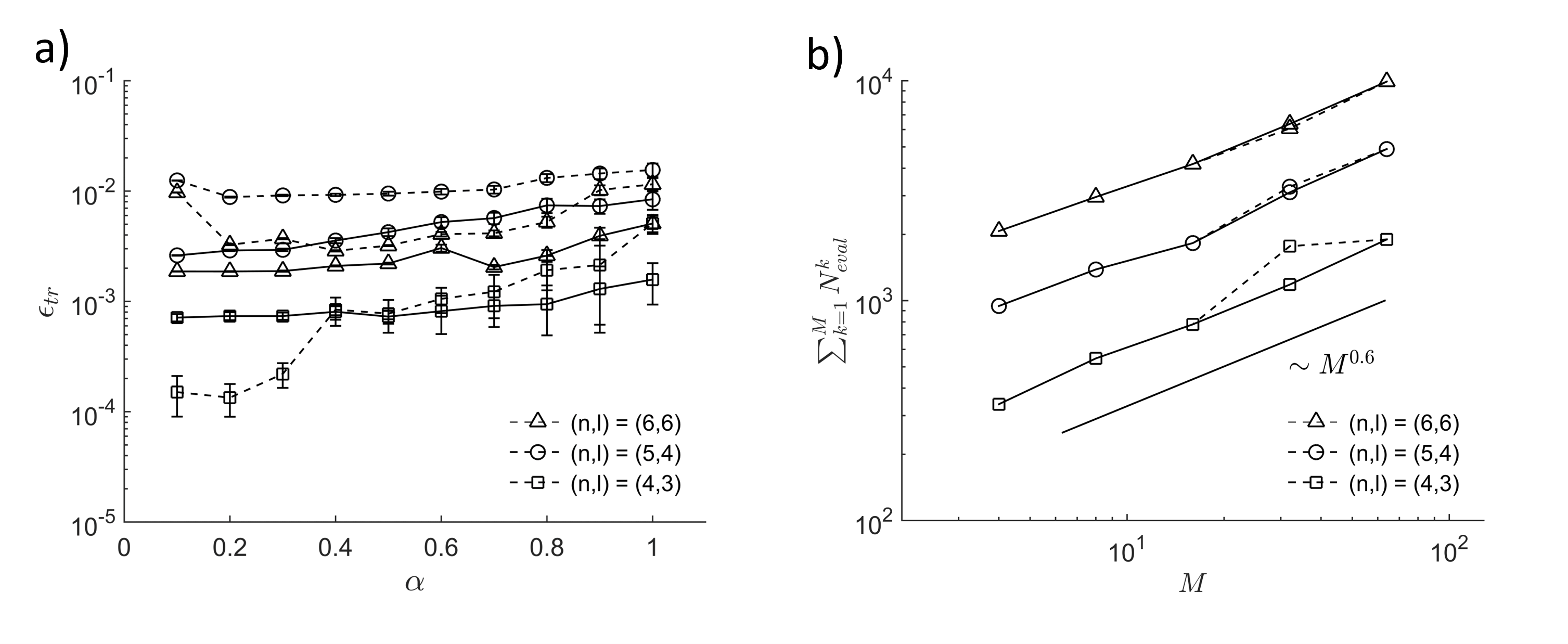}
    \caption{ (a) Time-averaged trace error $\epsilon_{\mathrm{tr}}$ plots for linear positive-amplitude hardware-efficient ansatz \cref{eq:HEAnsatz} (dotted lines) vs.\ circular ansatz \cref{eq:circular_ansatz} (continuous lines), for $0 < \alpha \leq 1$. Error bars denote standard deviation. (b) Number of function evaluations $\sum_{k=1}^M N_{eval}^k$ required by L-BFGS-B gradient-free optimizer for $\alpha = 0.5$ (dashed lines) and  $\alpha = 1$ (continuous lines), scales with number of time steps $M$ to a fractional power of $0.6$ in time $t=[0,0.5]$. Indicated $(n,l)$ pairs are selected based on the minimum number of layers $l$ required to encode $\ket{u(\boldsymbol{\theta}^0)}$ for $n\in\{3,4,5\}$.}
    \label{fig:fig3}
\end{figure}

Figure \ref{fig:fig3}a compares the results between the linear real-amplitude hardware-efficient ansatz \cref{eq:HEAnsatz} and the circular ansatz \cref{eq:circular_ansatz}, suggesting an improved solution fidelity for the circular ansatz over the linear ansatz ($n > 4$). According to unitary dependence theory \cite{Hu2023}, the effect of an additional circular entangling \textsc{cnot} gate ($C_nX_0$) on a single layer ansatz ($l=1$) is to transform the unitary dependence of the first qubit from $q_1:\{R_Y(\theta_1)\}$ to $q_1:\{R_Y(\theta_2) \sim R_Y(\theta_n)\}$, which effectively increases the connectivity of $q_1$ by $n-2$ qubits. This may improve the performance of an ansatz for certain problems~\cite{Hu2023}.

Figure \ref{fig:fig3}b shows that the number of function evaluations $\sum_{k=1}^M N_{eval}^k$, required by the L-BFGS-B gradient-free optimizer scales sub-linearly with the number of time steps $\mathcal{O} (M^{0.6})$ in time $t=[0,0.5]$. The overhead costs for classical gradient estimation remains economical up to $M = 2^6$.

\begin{figure}
    \centering
    \includegraphics[width=1.0\linewidth]{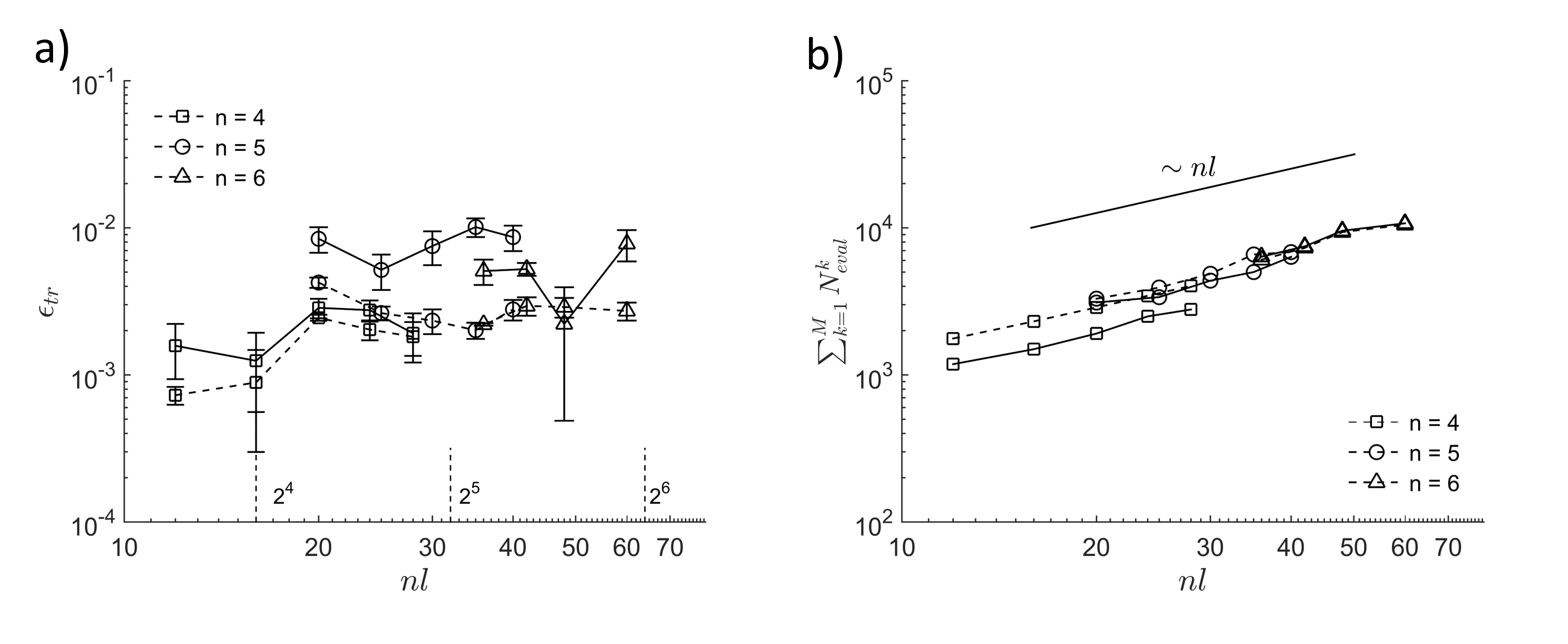}
    \caption{ (a) Time-averaged trace error $\epsilon_{\mathrm{tr}}$ is insensitive to over-parameterization, shown here for $\alpha = 0.5$ (dashed lines) and  $\alpha = 1$ (continuous lines). Error bars denote standard deviation. (b) Number of function evaluations $\sum_{k=1}^M N_{eval}^k$ scales linearly with number of parameters $nl$.}
    \label{fig:fig4}
\end{figure}

Figure \ref{fig:fig4} shows that increasing the number of parameters $nl$ does not improve the time-averaged trace error $\epsilon_{\mathrm{tr}}$, but instead increases the number of function evaluations which scales as $\mathcal{O} (nl)$. The linear scaling suggests that the problem of vanishing gradients due to excessive parameterization, commonly known as barren plateaus \cite{Cerezo2021}, may be mitigated by short time-steps, such that the initial solution at each time step is closer to the optimal solution.

\subsection{Time-fractional Burgers' equation} \label{subsec:burgers}
The fractional Burgers' equation \cite{Xu2013,Sugimoto1991} has found applications in shallow water waves and waves in bubbly liquids \cite{Garra2011}. We consider the following 1D time-fractional Burgers' equation \cite{Mukundan2016} as an extension to the time-fractional diffusion equation \cref{eq:1},
\begin{align}
    D_t^\alpha u(t,x) = \nu \partial_{xx} u(t,x) - u(t,x)\partial_{x} u(t,x),\quad 0 < \alpha \leq 1,
    \label{eq:burgers}
\end{align}
where $D_t^\alpha$ is the Caputo fractional derivative of the $\alpha$-th order, as previously defined, and $\nu$ is the fluid kinematic viscosity. Readers interested in quantum computing algorithms for non-fractional Burgers' equation ($\alpha = 1$) are referred to \cite{Liu2021,Oz2022,Lubasch2020}.

\subsubsection{Semi-implicit scheme}
Using explicit source terms, we apply central difference on the non-linear advection term, such that
\begin{align}
    \left( 1 - a \delta \right) u_{n}^k =
    w_k u_n^0 - \sum_{j=1}^{k-1} \Delta w_{j} u_n^{k-j}
    - b \left( u_{n+1}^{k-1} - u_{n-1}^{k-1} \right) u_{n}^{k-1},
\end{align}
where $a = \nu/(g_{\alpha,\tau} h^2)$ and $b = 1/(2g_{\alpha,\tau} h)$. In Dirac notation, the time-fractional Burgers' equation can be iterated as \cref{eq:fde} 
\begin{align}
    A \ket{\tilde{u}^k} = \ket{\tilde{f}^{k-1}} - b \left( \tilde{\Lambda}_{+}^{k-1} - \tilde{\Lambda}_{-}^{k-1} \right) \ket{\tilde{u}^{k-1}},
\end{align}
where $\tilde{\Lambda} := \mathrm{diag}(u)$ is a diagonal matrix with $u$ along its diagonal, and the subscript $\pm$ denotes either incremental or decremental shift in each eigenstate under periodic boundary conditions, i.e. $\ket{2^n}=\ket{0}$. 

With that, the cost function for Burgers' equation is
\begin{align}
    \mathcal{C}^k (r^k,\boldsymbol{\theta}^k) = 
    -\frac{1}{2} \frac{\left[ \langle u^k, \tilde{f}^{k-1} \rangle - b \left( \langle u^k,\tilde{\Lambda}_{+}^{k-1} \tilde{u}^{k-1}\rangle - \langle u^k,\tilde{\Lambda}_{-}^{k-1} \tilde{u}^{k-1}\rangle \right) \right]^2} 
    {\bra{u^k}A\ket{u^k}},
    \label{eq:burger_cost}
\end{align}
which includes two non-linear terms $\langle u^k,\tilde{\Lambda}_{\pm}^{k-1} \tilde{u}^{k-1}\rangle$ evaluated using redundant copies of $k-1$ quantum states  \cite{Lubasch2020, sarma2023quantum} on quantum circuit (\cref{fig:figA2}c).

\subsubsection{Test case}
We test a 1D bi-directional flow using time-fractional Burgers' equation \cref{eq:burgers} with initial condition $u(0,x) = \sin(2\pi x)$ and Dirichlet boundary condition $u(0,x) = x(1-x)$ and $u(t,0) = u(t,1) = 0$.

\begin{figure}
    \centering
    \includegraphics[width=1.0\linewidth]{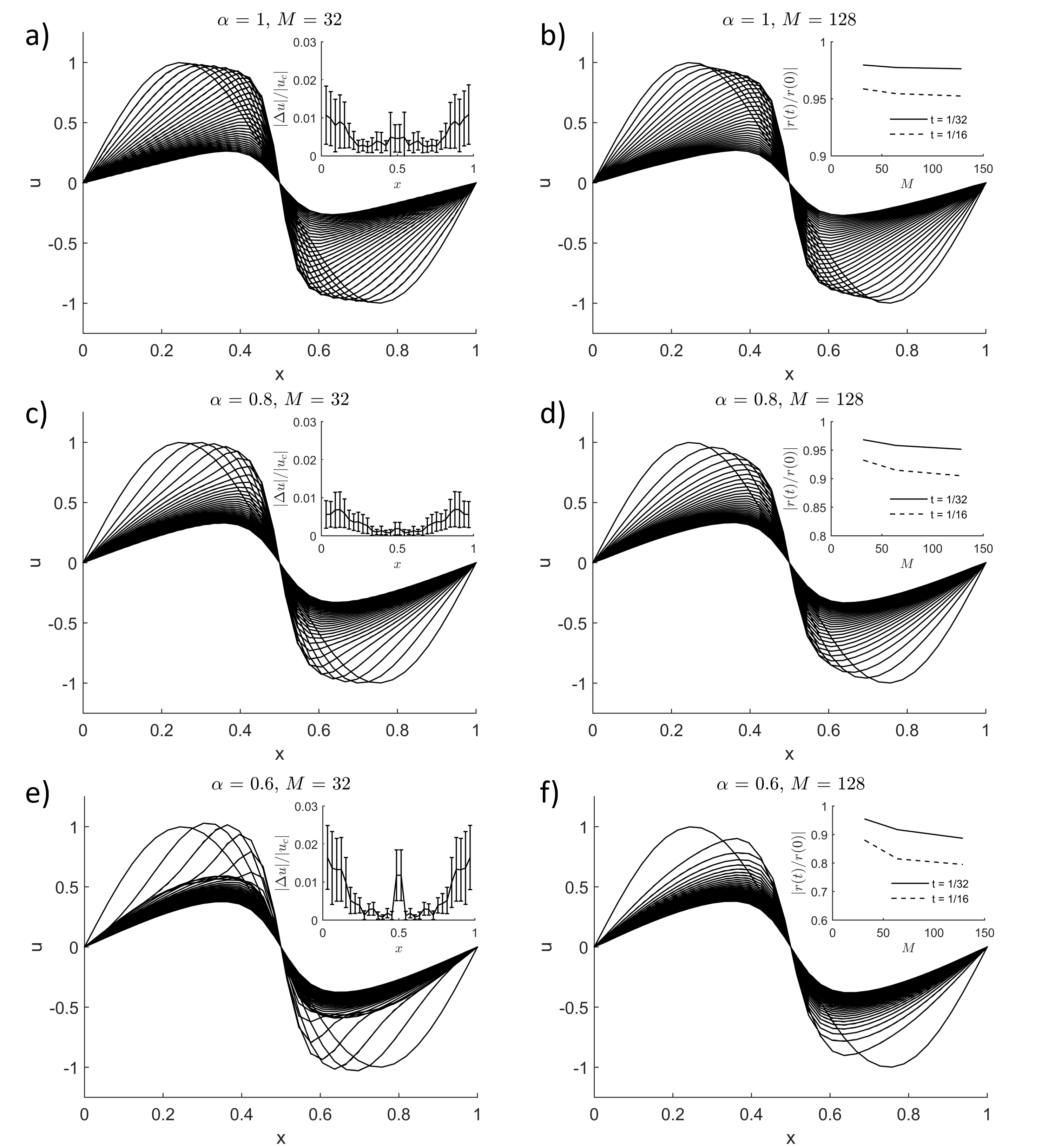}
    \caption{ Time-series solutions of 1D Burgers' equation plotted in time intervals of $1/32$ in (a,c,e) $N \times M = 32 \times 32$ and (b,d,f) $N \times M = 32 \times 128$ space-time domain of size $(L,T)=(1,1)$ for qubit-layer count $(n,l)= (5,5)$, kinematic viscosity $\nu = 0.02$ ($Re = 100$), and fractional order $\alpha$ (a,b) 1, (c,d) 0.8, and (e,f) 0.6. Insets (a,c,e) show time-averaged relative deviation from classical finite difference solution on the same domain and insets (b,d,f) show relative norm convergence with $M$ at times $t \in \{1/32, 1/16\}$. Initial and boundary conditions are $u(0,x) = \sin{(2\pi x)}$ and $u(t,0) = u(t,1) = 0$.}
    \label{fig:fig5} 
\end{figure}

\cref{fig:fig5}(a,c,e) show time series solutions in $32 \times 32$ space-time domain of size $(L,T)=(1,1)$ with qubit-layer count $(n,l)=(5,5)$ for fractional order $\alpha = \{1, 0.8, 0.6\}$ and kinematic viscosity $\nu = 0.02$. The Reynolds number is $Re \equiv 2u_c L/\nu = 100$, where the characteristic velocity $u_c = 1$. In the inviscid case, an initial sine-wave profile tends towards an N-wave shock solution. As with \cref{fig:fig2}, reducing fractional order leads to fast decay at short times and slow convergence at long times \cite{Akram2020}, shown here for $\alpha \in \{1,0.8,0.6\}$. Insets show the relative deviations from classical finite difference solution are less than 2\%, without any clear dependence on $\alpha$.

For semi-implicit time-fractional Burgers' equation, the Courant number $C \sim u_c \tau^{\alpha}/h$ depends on the fractional index $\alpha$. To verify the dependence of numerical stability on $t^{\alpha}$, the number of time steps $M$ is quadrupled from 32 to 128 in the time series solutions \cref{fig:fig5}(b,d,f). By inspection, time-fractional solutions are sensitive to time-step $\tau$. Specifically, for $\alpha = 0.6$, convergence in solution accuracy requires $M \gg N$ (see \cref{fig:fig5}f inset). 

\subsubsection{Trace error and optimization}

\begin{figure}
    \centering
    \includegraphics[width=1.0\linewidth]{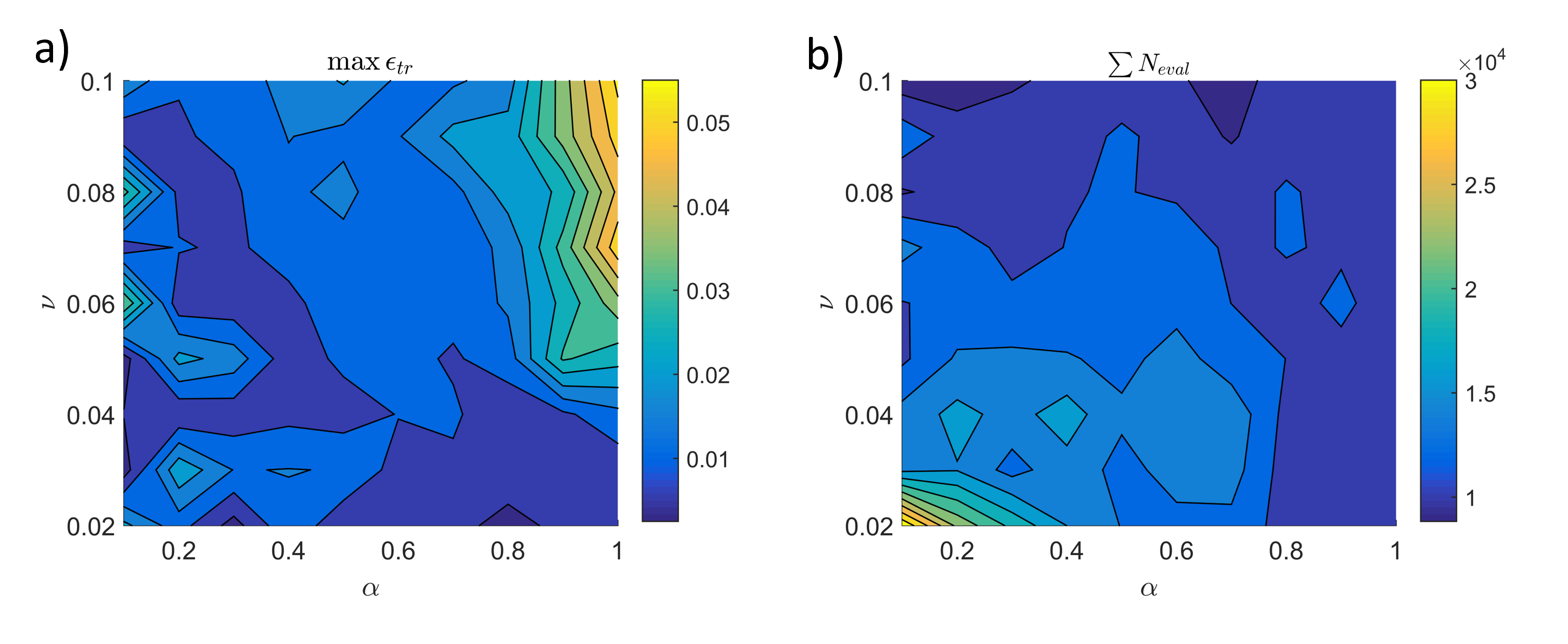}
    \caption{$\alpha$-$\nu$ contour plots of (a) maximum trace error $\max \epsilon_\mathrm{tr}$ and (b) number of function evaluations $\sum N_\mathrm{eval}$ for variational quantum solutions of 1D Burgers' equation (same as \cref{fig:fig5}).}
    \label{fig:fig6}
\end{figure}

\Cref{fig:fig6}a shows for $\nu \gtrsim 0.04$, the maximum trace error $\max \epsilon_\mathrm{tr}$ for normal Burgers' equation ($\alpha = 1$) is significantly greater than for the time-fractional Burgers' equation ($\alpha < 1$), the latter which converges to steady state solutions faster (\cref{fig:fig5}). \Cref{fig:fig6}b shows the required number of function evaluations $\sum N_\mathrm{eval}$ scales with both $\nu$ and $\alpha$, due to the difficulty of convergence towards unstable solutions driven by the explicit advection term. 

Together, these results suggest a trade-off between solution fidelity and gradient evaluation costs, depending on fractional index $\alpha$ and diffusion parameter $\nu$.

\subsection{Fractional diffusive epidemic model}\label{subsec:epidemic}

A natural extension to the nonlinear differential equations involves exploring coupled systems of differential equations, as exemplified by the susceptible-exposed-infectious-recovered (SEIR) epidemic model \cite{Liu2021}. Fractional-order epidemic models have garnered significant attention in the wake of the COVID-19 pandemic \cite{Ali2022,Rezapour2020,Khan2020}. These models include spatio-temporal models that account for geographical spread of the disease \cite{Barnes2023, Nazia2022, Hu2021}. A minimal representation of a fractional diffusive SEIR model is given by \cite[Eq.~(24)--(29)]{Alaje2023}:
\begin{align}
    D_t^{\alpha_1} S &= \nu_1 \partial_{xx} S - \beta IS - \mu S + \Pi, \label{eq:SEIR_S} \\
    D_t^{\alpha_2} E &= \nu_2 \partial_{xx} E + \beta IS - (\sigma + \mu) E, \label{eq:SEIR_E} \\
    D_t^{\alpha_3} I &= \nu_3 \partial_{xx} I + \sigma E - (\rho + \mu) I, \label{eq:SEIR_I} \\
    D_t^{\alpha_4} R &= \nu_4 \partial_{xx} R + \rho I - \mu R, \label{eq:SEIR_R}
\end{align}
which partitions a population into four distinct cohorts: susceptible individuals $S$, exposed individuals $E$, infectious individuals $I$, and those who have recovered $R$, without accounting for vaccination and quarantine. The symbols $\alpha_i$ and $\nu_i$, where $i\in \{1,2,3,4\}$, are the fractional order and the diffusion coefficient corresponding to each respective $\{S,E,I,R\}$ cohort in that order, $\Pi$ is the population influx rate, $\beta$ is the infective rate, $\mu$ is the death rate, $\sigma$ is the progression rate and $\rho$ is the recovery rate.                       

\subsubsection{Coupled semi-implicit scheme}
The coupled system of \cref{eq:SEIR_S,eq:SEIR_E,eq:SEIR_I,eq:SEIR_R} is discretized via finite difference and expressed in the following iterative form with common time-step $\tau$,
\begin{align}
    \left(\beta I^{k-1} + \mu + g_{\alpha_1,\tau} - \frac{\nu_1}{h^2} \delta \right) S^k &=
    g_{\alpha_1,\tau} f_{\alpha_1} \left(S^{[0,k-1]}\right) + \Pi, \label{eq:SEIR_Sk} \\
    \left( \sigma + \mu + g_{\alpha_2,\tau} - \frac{\nu_2}{h^2} \delta \right) E^k &= g_{\alpha_2,\tau} f_{\alpha_2} \left(E^{[0,k-1]}\right) + \beta I^{k-1}S^{k-1}, \label{eq:SEIR_Ek} \\
    \left( \rho + \mu + g_{\alpha_3,\tau} - \frac{\nu_3}{h^2} \delta \right) I^k &= g_{\alpha_3,\tau} f_{\alpha_3} \left(I^{[0,k-1]}\right) + \sigma E^{k-1}, \label{eq:SEIR_Ik} \\
    \left( \mu + g_{\alpha_4,\tau} - \frac{\nu_4}{h^2} \delta \right) R^k &= g_{\alpha_4,\tau} f_{\alpha_4} \left(R^{[0,k-1]}\right) + \rho I^{k-1}, \label{eq:SEIR_Rk}
\end{align}
where the operator $f_{\alpha} \left(u^{[0,k-1]}\right) =  w_k^{[\alpha]} u^0 - \sum_{j=1}^{k-1} \Delta w_{j}^{[\alpha]} u^{k-j}$. Since the $\beta IS$ term is strictly positive, it can be iterated as $\beta I^{k-1}S^{k}$ (\cref{eq:SEIR_Sk}). With only an $\mathcal{O}(1)$ number of cohorts, the coupled system of equations can be solved using a separate cost function for each cohort at time $k$, i.e.
\begin{align}
    \mathcal{C}^k_S &=
    \frac{(r^k_S)^2}{2} \left( \bra{S^k}A_S\ket{S^k} + \beta \bra{S^k}\tilde{\Lambda}^{k-1}_I\ket{S^k} \right)
    - r^k_S \left( g_{\alpha_1,\tau} \langle S^k,\tilde{f}_{S,\alpha_1}^{k-1} \rangle  + \Pi \langle S^k,\tilde{+}^n \rangle \right), \label{eq:SEIR_CS} \\
    \mathcal{C}^k_E &=
    \frac{(r^k_E)^2}{2} \bra{E^k}A_E\ket{E^k}
    - r^k_E \left( g_{\alpha_2,\tau} \langle E^k,\tilde{f}_{E,\alpha_2}^{k-1} \rangle  + \beta \langle E^k,\tilde{\Lambda}^{k-1}_I \tilde{S}^{k-1} \rangle \right), \label{eq:SEIR_CE} \\
    \mathcal{C}^k_I &=
    \frac{(r^k_I)^2}{2} \bra{I^k}A_I\ket{I^k}
    - r^k_I \left( g_{\alpha_3,\tau} \langle I^k,\tilde{f}_{I, \alpha_3}^{k-1} \rangle  + \sigma \langle I^k, \tilde{E}^{k-1} \rangle \right), \label{eq:SEIR_CI} \\
    \mathcal{C}^k_R &=
    \frac{(r^k_R)^2}{2} \bra{R^k}A_R\ket{R^k}
    - r^k_R \left( g_{\alpha_4,\tau} \langle R^k,\tilde{f}_{R, \alpha_4}^{k-1} \rangle  + \rho \langle R^k, \tilde{I}^{k-1} \rangle \right), \label{eq:SEIR_CR} 
\end{align}
where $|\tilde{+}^n\rangle \equiv \sqrt{2^n}\ket +^{\otimes n}$ is the (unnormalized) equal superposition over all computational basis states, and each subscript label $\mathbb{S} \in \{S,E,I,R\}$ corresponds to the indicated cohort. The Hamiltonian matrices $\in \mathbb{R}^{N \times N}$  are $A_S = (\mu + g_{\alpha_1,\tau})\mathcal{I}-\nu_1/h^2 \mathcal{L}$, $A_E = (\sigma + \mu + g_{\alpha_2,\tau})\mathcal{I}-\nu_2/h^2 \mathcal{L}$, $A_I = (\rho + \mu + g_{\alpha_3,\tau})\mathcal{I}-\nu_3/h^2 \mathcal{L}$ and $A_R = (\mu + g_{\alpha_4,\tau})\mathcal{I}-\nu_4/h^2 \mathcal{L}$, where $\mathcal{L}$ is a symmetric tridiagonal matrix with $-2$ along its main diagonal and $1$ along the adjacent off-diagonals, and $\mathcal{I}=\mathbb{I}^{\otimes n}$ is an $n$-qubit identity matrix. Terms such as $\bra{S(\boldsymbol{\theta}^k)}\Lambda^{k-1}_I \ket{S(\boldsymbol{\theta}^k)}$ can be measured using the circuit shown in \cref{fig:figA2}d. 

\subsubsection{Basic reproduction number}
The spread of a transmissible disease is governed by an important quantity known as the basic reproduction number \cite{VandenDriessche2017},
\begin{align}
    \mathcal{R} = \frac{\beta \sigma \Pi}{\mu(\mu+\sigma)(\mu+\rho)},
\end{align}
which is defined, in epidemiological terms, as the average number of secondary cases produced by one infected individual in a susceptible population. Depending on the value of $\mathcal{R} $, the number of infectious individuals may increase resulting in an epidemic ($\mathcal{R}>1$), or decrease resulting in the disease dying out ($\mathcal{R}<1$).

\subsubsection{Test case}
Consider a set of $2^n$ spatially connected populations, each with a small infected cohort and a much larger susceptible cohort. Following \cite{Kammegne2023}, we assume the following set of SEIR parameters: $\Pi = 750$ \cite{Alaje2023}, $\mu = 0.03325$, $\beta = 5.1 \times 10^{-6}$, $\sigma = 0.17$, and $\rho = 0.1109$, leading to $\mathcal{R} \approx 0.66$. Individuals are assumed to move around within a region $[0,1]$ according to Fick's law, with diffusion coefficents $\nu_1=\nu_2=10^{-3}$ and  $\nu_3=\nu_4=5 \times 10^{-4}$. Initial cohort sizes are $S(0,x) = 22,500$ ($\sim \Pi/\mu$ for near-steady populations \cite{Kammegne2023}), $I(0,x) = 20 e^{-2x}$ (for spatial dependence \cite{Alaje2023}) and $E(0,x) = R(0,x) = 1$. Neumann boundary conditions are applied, i.e. $\partial_x \mathbb{S}(t,0) = \partial_x \mathbb{S}(t,L) = 0$, where $\mathbb{S} \in \{S,E,I,R\}$.

\begin{figure}
\centering
\includegraphics[width=1.0\linewidth]{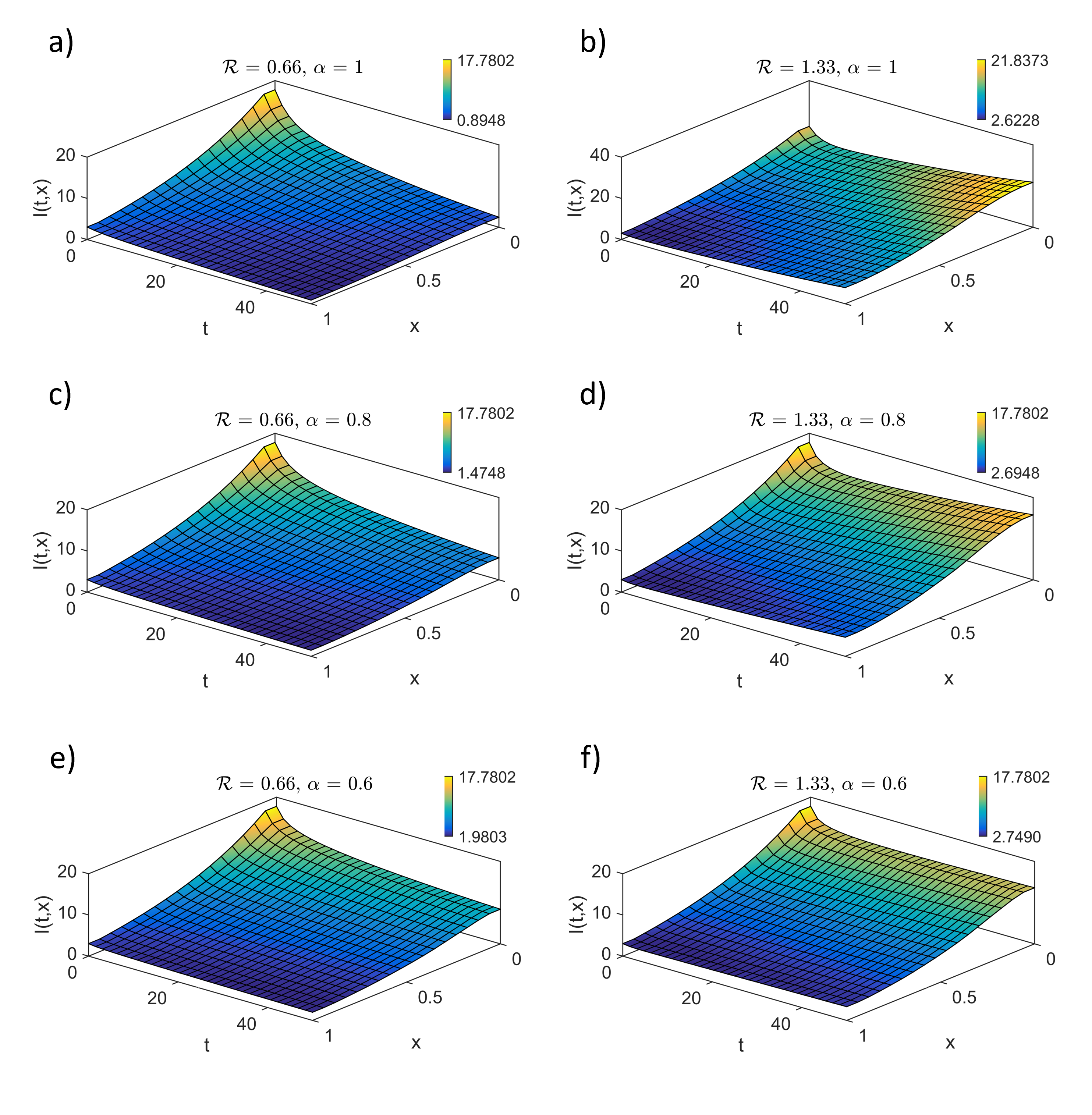}
\caption{Infectious $I$ cohort (actual values) in $16 \times 32$ space-time domain of size $(L,T)=(1,100)$ with qubit-layer count $(n,l)=(4,5)$ for each cohort, depending on basic reproduction number (a,c,e) $\mathcal{R} = 0.66$ and (b,d,f) $\mathcal{R} = 1.33$, and fractional index (a,b) $\alpha = 1$, (c,d) $\alpha = 0.8$ and (e,f) $\alpha = 0.6$.} 
\label{fig:fig7}
\end{figure}

In a $16 \times 32$ domain of size $(L,T) = (1,100)$ with $(n,l) = (4,5)$, \cref{fig:fig7} verifies that $\mathcal{R}<1$ results in a monotonic decrease (a,c,e) in the infectious cohort $I$, whereas $\mathcal{R}>1$ results in an increase (b,d,f) in $I$ before eventual decrease at long times (not shown). A decrease in the fractional index $\alpha$ flattens the spatiotemporal profile of $I$, in such a way that $I$ decreases more slowly for $\mathcal{R}<1$, and increases more slowly for $\mathcal{R}>1$.

\begin{figure}
\centering
\includegraphics[width=1.0\linewidth]{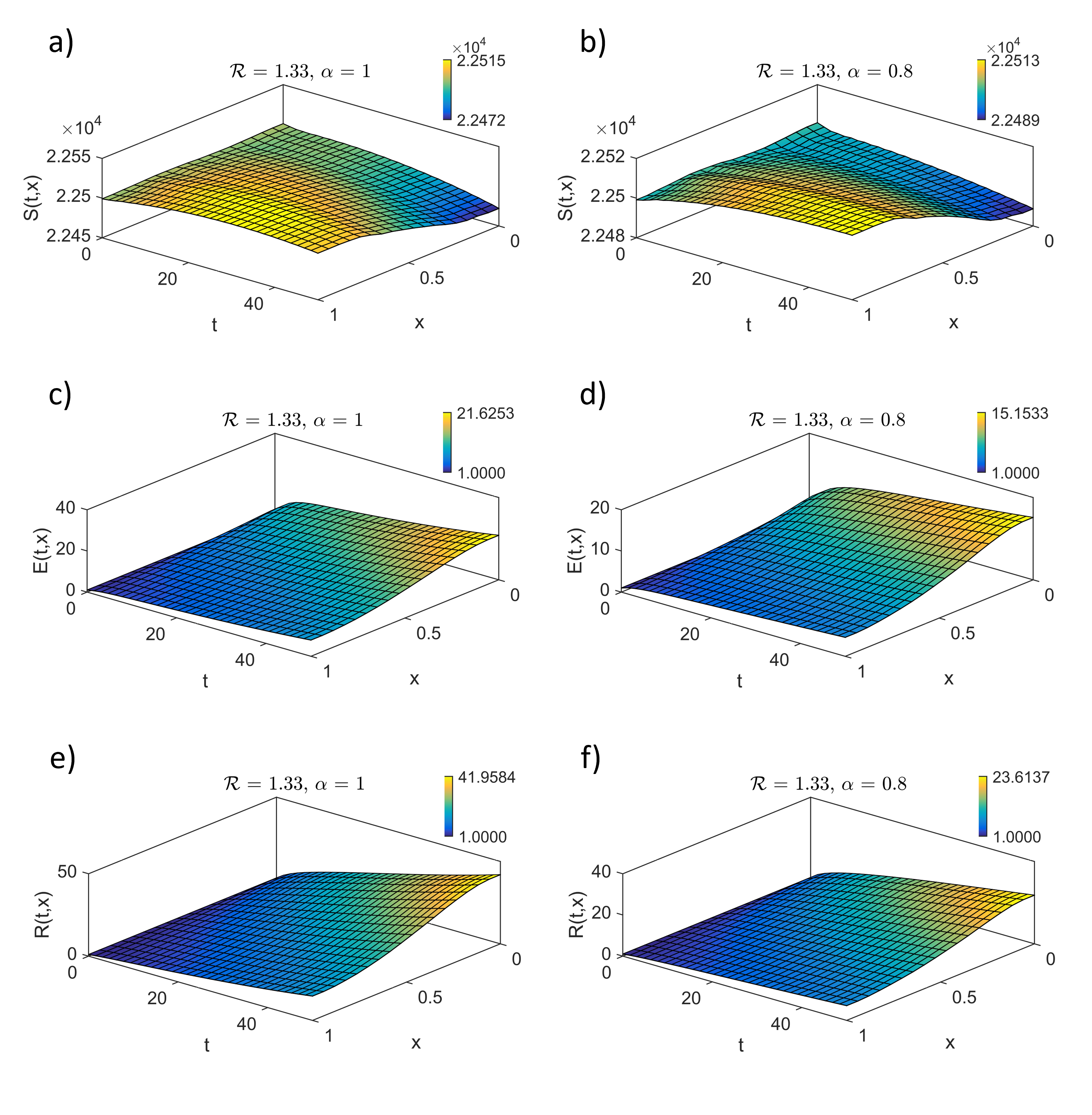}
\caption{(a,b) Susceptible $S$, (c,d) exposed $E$ and (e,f) recovered $R$ cohorts (actual values) in $16 \times 32$ space-time domain of size $(L,T)=(1,100)$ with qubit-layer count $(n,l)=(4,5)$ for each cohort at the start of an epidemic ($\mathcal{R} = 1.33 > 1$). Fractional index (a,c,e) $\alpha = 1$ and (b,d,f) $\alpha = 0.8$.
} 
\label{fig:fig8}
\end{figure}

Similar time-fractional memory effects also apply to other cohorts, namely susceptible $S$, exposed $E$ and recovered $R$, as shown in \cref{fig:fig8} at the start of an epidemic ($\mathcal{R} > 1$). The apparent smoothness of the solutions, as presented, affirms the potential of the iterative solver in other applications involving coupled fractional-order derivatives, such as predator-prey population models \cite{Jafari2021} and  Oldroyd-B viscoelastic flows \cite{Wang2023}.

\section{Hardware noise}\label{sec:hardware}

Near-term quantum devices are subjected to hardware noise which significantly impairs the performance of variational quantum algorithms \cite{Pellow-Jarman2021, Bonet-Monroig2023}, despite their inherent resilience to coherent errors \cite{Fontana2020}. To explore the effect of noise, we employ a realistic noise model sampled from the IBM Nairobi quantum hardware device (Falcon r5.11H, version 1.3.3) applied to the \emph{Aer Simulator} backend provided by the \emph{Pennylane-Qiskit} plugin. Circuits are transpiled using default settings with level 3 optimization and SABRE routing. The optimizer of choice is the Simultaneous Perturbation Stochastic Approximation (SPSA), which is known for its exceptional scalability and performance under noisy conditions \cite{Pellow-Jarman2021, Bonet-Monroig2023}. We apply the SPSAOptimizer class on \emph{Pennylane} using default hyperparameters for 200 iterations, equivalent to 400 circuit evaluations. Data readout is assumed to be noiseless via statevector emulation.

\begin{figure}
\centering
\includegraphics[width=1.0\linewidth]{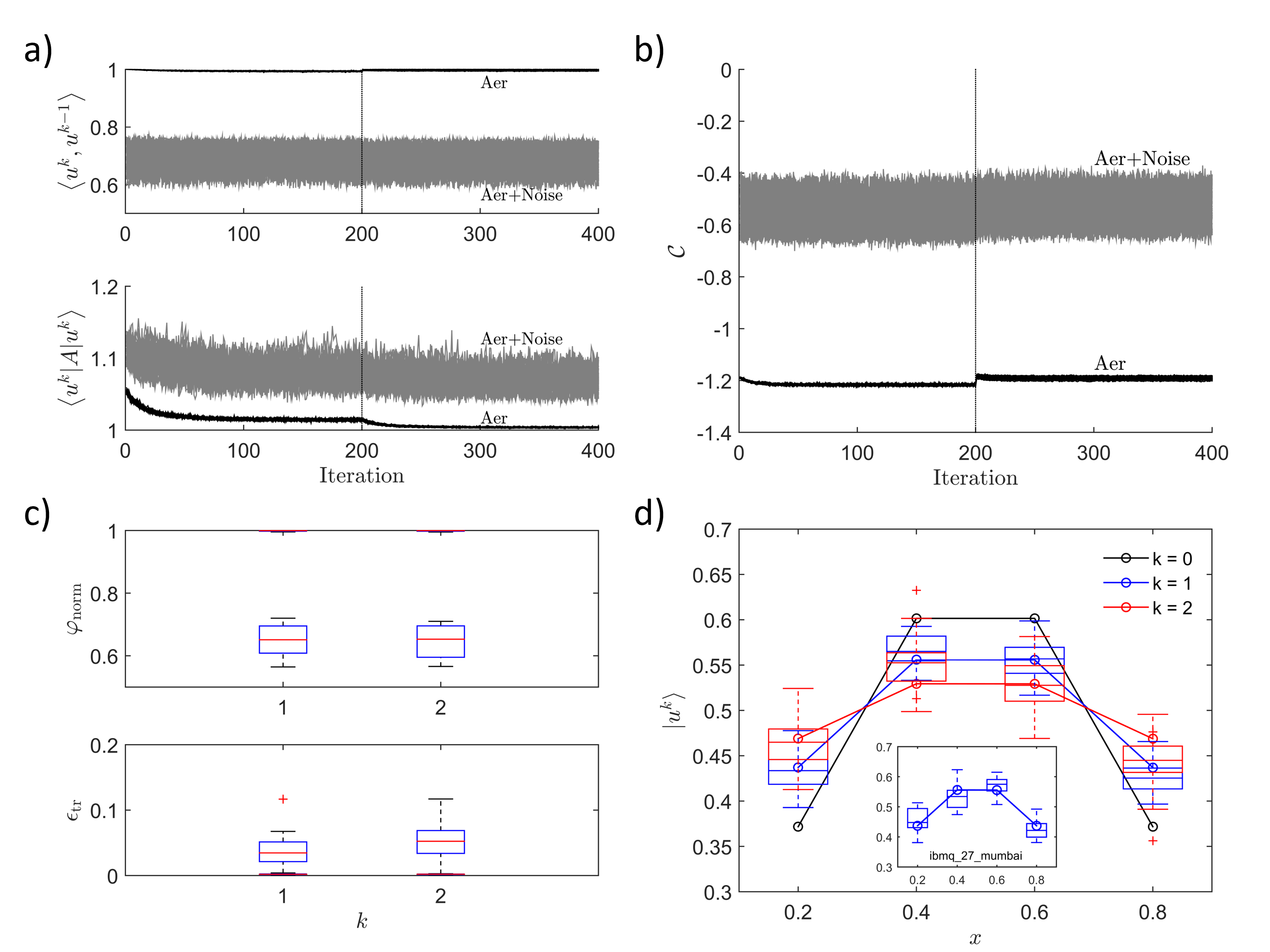}
\caption{ Hardware noise model tests across 40 instances on \emph{Aer Simulator} using SPSA optimizer set at 200 iterations per time step. (a) Effect of noise model on overlap and Hamiltonian measurements results in shifting of (b) the optimized cost function $\mathcal{C}$. (c) Box plots of norm fidelity $\varphi_{\mathrm{norm}}$ and trace error $\epsilon_{\mathrm{tr}}$ with noise model in time steps $k = \{1,2\}$. Reference values without noise model are shown as $\varphi_{\mathrm{norm}} \approx 1$ and $\epsilon_{\mathrm{tr}} \approx 0$. (d) Comparing $\ket{u^{(1)}}$ and $\ket{u^{(2)}}$ with noise model (box plots) to noiseless simulation (line circles). Inset shows hardware test results across 20 instances obtained from the IBMQ Mumbai 27 qubit device using Qiskit Runtime set at 200 SPSA iterations for a single time step.
}
\label{fig:fig9}
\end{figure}

Here, we solve minimally for the sub-diffusion problem (Section \ref{subsec:subdiffusion}) in $N \times M = 4 \times 2$ using qubit-layer count $(n,l)=(2,1)$ for $\alpha = 1$. Initial conditions are $u(0,x) = x(1-x)$ and boundary conditions are periodic instead of Dirichlet. \Cref{fig:fig9} shows the effects of hardware noise model on \emph{Aer Simulator} based on 40 instances with fixed 10,000 shots per circuit evaluation. Of note here is the effect of noise on the evaluation of the norm, which directly affects the weights of the cost function $\mathcal{C}$ for the subsequent time step. For fair evaluation of each time-step, we restore the norm to the classical error-free value before each time-step.

\Cref{fig:fig9}a shows that averaged overlap measurements are shifted by the noise model to $\langle u^{1}, u^{0} \rangle = 0.686 \pm 0.049$ and $\langle u^{2}, u^{1} \rangle = 0.682 \pm 0.049$, from approximately $ 1$ for measurements without noise model. The observed relative noise error is approximately $ 30\%$, which is significantly greater than the error from Hamiltonian measurements $\bra{u^k}A\ket{u^k}$ with noise model (less than $10\%$ by inspection, \cref{fig:fig9}b). This can be attributed to greater gate errors from the Toffoli (controlled \textsc{cnot}) gates found in controlled entangling layers of the overlap circuit. In addition, the fractional noise error of the cost function \cref{eq:cost_function} 
\begin{align}
    \left(\eta_{C}\right)^2 = \left(2\eta_{O}\right)^2 + \left(\eta_{H}\right)^2
\end{align}
depends on $\eta_{O}$, the fractional noise error of the overlap measurements $\langle u^k, u^{k-1} \rangle$, and $\eta_{H}$, the fractional noise error of the Hamiltonian measurements $\bra{u^k}A\ket{u^k}$. Given $\eta_{O}/\eta_{H} \approx 3$, the proportion of noise error of the cost function due to the noise of overlap measurement is $\sqrt{36/37} \approx 0.986$. This means that the overall performance of the algorithm depends almost entirely on the noise error of overlap measurements, and very little on that of Hamiltonian measurements.

\Cref{fig:fig9}c shows the norm fidelity 
\begin{align}
    \varphi_{\mathrm{norm}}^k = \left| \frac{r^k}{\|u^k\|} \right|,
\end{align}
where $r^k$ is the measured norm \cref{eq:norm} and $\|u^k\|$ is the classical norm, and trace error $\epsilon_{\mathrm{tr}}^k$ \cref{eq:trace_error} at time step $k\in\{1,2\}$. Here, the mean norm fidelities $\overline{\varphi}_{\mathrm{norm}}^k \approx 0.65$ are nearly identical for $k\in\{1,2\}$ due to the classical norm reset, whereas the mean trace errors are $\overline{\epsilon}_{\mathrm{tr}}^{1} \approx 0.039$ and $\overline{\epsilon}_{\mathrm{tr}}^{2} \approx 0.052$. 

\Cref{fig:fig9}d shows the effects of hardware noise on the quantum solution states $\ket{u^k}$ over 40 instances. Although the degree of data scatter at each node (width of box plot) is of a similar magnitude to the difference between consecutive classical solutions, the mean readout states $\ket{\overline{u}^k}$ are in general agreement with the respective classical solutions. Further tests were conducted on actual hardware, in this case, the IBMQ Mumbai 27 qubit device across 20 instances (see inset) using Qiskit Runtime sampler and estimator primitives for overlap and Hamiltonian measurements respectively, set at 200 SPSA iterations for a single time step.

\section{Discussion}\label{sec:discussion}

In this study, we proposed and implemented a hybrid variational quantum algorithm for solving time-fractional diffusive differential equations in engineering applications, including non-linear equations such as Burgers' equation (Section \ref{subsec:burgers}) and coupled systems of equations such as those in epidemic models (Section \ref{subsec:epidemic}). Our approach is not only efficient in time complexity, but also offers advantages in terms of memory utilization. While the classical memory cost of numerically solving these equations scales linearly with the number of spatial grid points $N$, the quantum memory cost scales only logarithmically with $N$. This helps to mitigate the global dependence problem in fractional calculus, where previous solutions must be repeatedly accessed from memory causing the classical solver to have prohibitive memory costs when the spatial range is large. In terms of time complexity, the present iterative scheme is limited by the quadratic scaling with the number of time steps; this has some implications in the numerical stability of non-linear fractional equations (see \cref{fig:fig5}).

The choice of the ansatz in the form of a parameterized circuit is reflected in its expressibility and entangling capability performance \cite{Sim2019} of a variational quantum algorithm \cite{cerezo2021variational}. Hence, as a matter of heuristics, the real-amplitude ansatze used in this study (\cref{eq:HEAnsatz,eq:circular_ansatz}) are by no means the most efficient designs for such applications. Designing and optimizing ansatz architecture, including utilizing adaptive methods \cite{Grimsley2019} or leveraging geometric tools \cite{katabarwa2022connecting,haug2021capacity,haug2023natural}, remains an active area of research \cite{Qin2023,you2021exploring,tilly2022variational}.

While the hybrid approach offers notable advantages, it is not exempt from the challenges that are commonly encountered in variational quantum algorithms. First, unlike a classical computer, the cost of loading initial conditions as classical data as an \emph{n}-qubit quantum state could be prohibitive \cite{Aaronson2015}. Encoding schemes, such as parametric optimization (\cref{eq:encoding}), may incur an exponential cost in primitive operations since the encoded \emph{n}-qubit state occupies a space of dimension $O(2^n)$. More efficient encoding techniques are currently being developed; for example, solutions may be represented in the form of finite Fourier series, using the quantum Fourier transform as part of the Fourier series loader introduced by \cite{Moosa2024} and employed by \cite{sarma2023quantum}. The use of the Walsh series instead of the Fourier series may yield an even more efficient sub-exponential scaling in encoding \cite{Zylberman2023}. In addition, extracting the complete solution at each of the $M$ time steps would require full readout of the statevector via quantum state tomography. This experimental procedure is notorious for its exponential cost \cite{o2016efficient}, which is a bottleneck to quantum advantage as pointed out by \cite{Aaronson2015}. Therefore, where possible, applications should consider a partial readout of the (spatial) solutions only at few selected time points of interest to minimize the expensive overhead due to tomography, or measuring only integrated quantities like mean, variance or other moments. In these cases, robust classical shadow tomography techniques \cite{huang2020predicting, bu2024classical, chen2021robust, koh2022classical, grier2022sample} can be employed to further reduce the required sample complexity.

In closing, the hybrid algorithm investigated in this work presents a nascent but promising pathway to solving fractional partial differential equations of the integro-differential form, subjected to potentially prohibitive space or memory requirements. More work needs to be done to address issues related to the trainability of variational quantum algorithms, such as barren plateaus and narrow gorges \cite{Cerezo2021, arrasmith2022equivalence}, as well as practical hardware implementation, including error mitigation \cite{cai2023quantum} and correction. 

\begin{appendices}

\section{Quantum circuit diagrams} \label{sec:A}

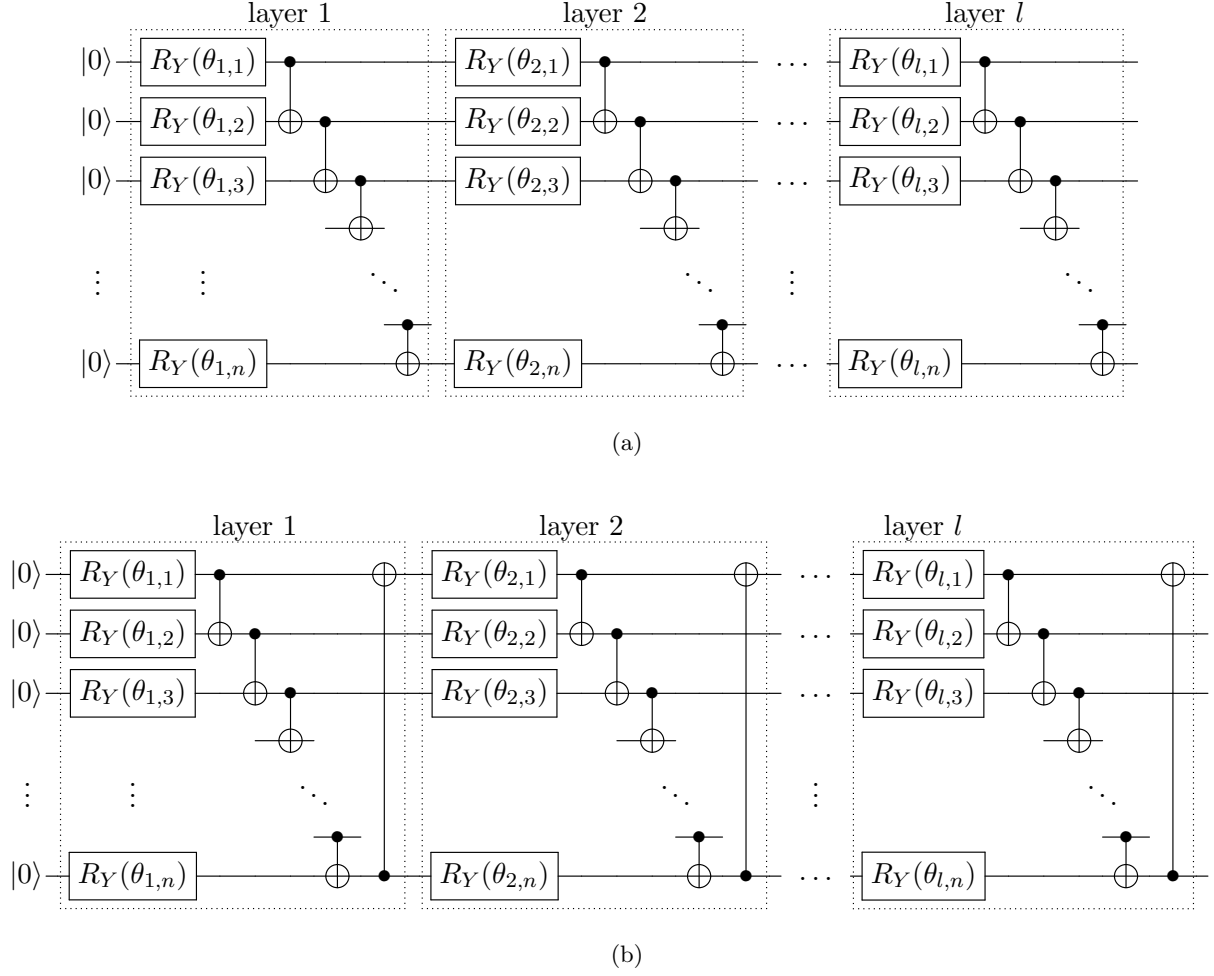
\begin{figure}[ht]
\centering
\begin{subfigure}[]{\textwidth}
\[ 
\Qcircuit @C=0.4em @R=.4em {
 &  & & \mbox{layer 1} 
& &  & & & & & & \mbox{layer 2} & & & & &  &  & & & & & & & &  \mbox{layer $l$}  & 
\\~\\
& \lstick{\ket{0}}\qw & \gate{R_Y(\theta_{1,1})} & \ctrl{1} & \qw  & \qw & \qw & \qw & \qw & \qw & \gate{R_Y(\theta_{2,1})} & \ctrl{1} & \qw  & \qw & \qw & \qw & \qw & \qw &  & &  \ldots & & & &  \gate{R_Y(\theta_{l,1})} & \ctrl{1} & \qw & \qw   & \qw & \qw & \qw & \qw
\\ & \lstick{\ket{0}}\qw & \gate{R_Y(\theta_{1,2})} & \targ    & \ctrl{1} & \qw   & \qw & \qw & \qw & \qw & \gate{R_Y(\theta_{2,2})} & \targ    & \ctrl{1} & \qw   & \qw & \qw & \qw & \qw &  & &  \ldots & &  & & \gate{R_Y(\theta_{l,2})} & \targ    & \ctrl{1} & \qw   & \qw & \qw & \qw & \qw
\\
& \lstick{\ket{0}}\qw & \gate{R_Y(\theta_{1,3})} & \qw & \targ & \ctrl{1} & \qw & \qw & \qw & \qw & \gate{R_Y(\theta_{2,3})} & \qw & \targ & \ctrl{1} & \qw & \qw & \qw & \qw  & & & \ldots & & & &   \gate{R_Y(\theta_{l,3})} & \qw      & \targ    &   \ctrl{1} & \qw & \qw & \qw & \qw 
\\
& &  & & & \targ & \qw &  & & & & & & \targ & \qw & & & & & & & & &  & & & & \targ & \qw & 
\\~\\~\\
\lstick{\vdots} & & \vdots & & &  &  \ddots  & & & & & & & & \ddots & & & & & &  \vdots & & & & & & & & \ddots &
\\~\\~\\~\\
& & & & & & &  \ctrl{1} & \qw & & & & & & & \ctrl{1} & \qw  & & & & & & & & & & & & & \ctrl{1} & \qw &
\\
& \lstick{\ket{0}}\qw & \gate{R_Y(\theta_{1,n})} & \qw & \qw & \qw & \qw & \targ & \qw & \qw & \gate{R_Y(\theta_{2,n})} & \qw & \qw & \qw & \qw & \targ & \qw & \qw  & & & \ldots & & & &   \gate{R_Y(\theta_{l,n})} & \qw  & \qw  & \qw  & \qw      & \targ    &     \qw & \qw
\gategroup{3}{3}{14}{8}{.6em}{.}
\gategroup{3}{11}{14}{16}{.6em}{.}
\gategroup{3}{25}{14}{30}{.6em}{.}
}
\]
\caption{}
\end{subfigure}

 \vspace{0.5cm}

\begin{subfigure}[]{\textwidth}
\[
\Qcircuit @C=0.4em @R=.4em {
 &  & & & \mbox{layer 1} 
& &  & &  & & & & & \mbox{layer 2} & & & & &  &  & & & & & & &  & & \mbox{layer $l$}  & 
\\~\\
& \lstick{\ket{0}}\qw & \gate{R_Y(\theta_{1,1})} & \ctrl{1} & \qw  & \qw & \qw & \qw & \qw &  \targ & \qw & \qw & \gate{R_Y(\theta_{2,1})} & \ctrl{1} & \qw  & \qw & \qw & \qw & \qw & \targ & \qw & \qw & & &  \ldots & & & &  \gate{R_Y(\theta_{l,1})} & \ctrl{1} & \qw & \qw   & \qw & \qw  & \qw & \targ & \qw & \qw
\\ & \lstick{\ket{0}}\qw & \gate{R_Y(\theta_{1,2})} & \targ    & \ctrl{1} & \qw   & \qw & \qw & \qw & \qw & \qw & \qw & \gate{R_Y(\theta_{2,2})} & \targ    & \ctrl{1} & \qw   & \qw & \qw & \qw & \qw &  \qw & \qw & & &  \ldots & &  & & \gate{R_Y(\theta_{l,2})} & \targ    & \ctrl{1} & \qw   & \qw & \qw & \qw & \qw & \qw & \qw
\\
& \lstick{\ket{0}}\qw & \gate{R_Y(\theta_{1,3})} & \qw & \targ & \ctrl{1} & \qw & \qw & \qw & \qw & \qw & \qw & \gate{R_Y(\theta_{2,3})} & \qw & \targ & \ctrl{1} & \qw & \qw & \qw & \qw  & \qw & \qw & & & \ldots & & & &   \gate{R_Y(\theta_{l,3})} & \qw      & \targ    &   \ctrl{1} & \qw & \qw & \qw & \qw & \qw & \qw
\\
& &  & & & \targ & \qw & & & & & & & & & \targ & \qw & & & & & & & & & & & & & & & \targ & \qw & 
\\~\\~\\
\lstick{\vdots} & & \vdots & & &  &  \ddots  & & & & & & & & & & \ddots & & & & & & & &  \vdots & & & & & & & & \ddots &
\\~\\~\\~\\
& & & & & & &  \ctrl{1} & \qw & & & & & & & & & \ctrl{1} & \qw  & & & & & & & & & & & & & & & \ctrl{1} & \qw &
\\
& \lstick{\ket{0}}\qw & \gate{R_Y(\theta_{1,n})} & \qw & \qw & \qw & \qw & \targ & \qw &  \ctrl{-11} & \qw & \qw & \gate{R_Y(\theta_{2,n})} & \qw & \qw & \qw & \qw & \targ & \qw & \ctrl{-11} & \qw  & \qw & & & \ldots & & & &   \gate{R_Y(\theta_{l,n})} & \qw  & \qw  & \qw  & \qw      & \targ    &     \qw  & \ctrl{-11} & \qw & \qw
\gategroup{3}{3}{14}{10}{.6em}{.}
\gategroup{3}{13}{14}{20}{.6em}{.}
\gategroup{3}{29}{14}{36}{.6em}{.}
}
\]
\caption{}
\end{subfigure}

\caption{ (a) Real-amplitude hardware-efficient ansatz (\cref{eq:HEAnsatz}) and (b) circular real-amplitude ansatz (\cref{eq:circular_ansatz}).}
\label{fig:figA1}
\end{figure}

\begin{figure}[ht]
\centering

\begin{subfigure}[]{\textwidth}
\[
\Qcircuit @C=0.3em @R=0.05em @! 
{
\lstick{\ket 0} & \gate{H} & \ctrlo{1} & \ctrl{1} & \gate{H} & \meter & \qw \\
\lstick{\ket 0^{\otimes n}} & \qw & \gate{U(\boldsymbol{\theta}^k)} & \gate{U^j} & \qw & \qw & \qw
}
\]
\caption{}
\end{subfigure}

\vspace{0.5cm}

\begin{subfigure}[]{\textwidth}
\[
\Qcircuit @C=1.4em @R=0.5em
{
& & & & & & \mathcal{H} \\
\lstick{\ket 0^{\otimes n}} & \qw & \gate{U(\boldsymbol{\theta}^k)} & \qw & \gate{\mathcal{I}/\mathcal{S}} & \qw & \meter & \qw 
}
\]
\caption{}
\end{subfigure}

\vspace{0.5cm}

\begin{subfigure}[]{\textwidth}
\[
\Qcircuit @C=0.3em @R=0.05em @! 
{
\lstick{\ket 0} & \gate{H} & \ctrlo{1} & \ctrl{2} & \ctrl{1} & \ctrl{1} & \gate{H} & \meter & \qw 
\\
\lstick{\ket 0^{\otimes n}} & \qw & \gate{U(\boldsymbol{\theta}^k)} & \qw & \gate{U^{k-1}} & \ctrl{1} & \qw & \qw & \qw 
\\
\lstick{\ket 0^{\otimes n}} & \qw & \qw & \gate{U^{k-1}} & \gate{\mathcal{S}/\mathcal{S}^{\dagger}} & \targ & \qw & \qw & \qw \\
}
\]
\caption{}
\end{subfigure}

 \vspace{0.5cm}

\begin{subfigure}[]{\textwidth}
\[
\Qcircuit @C=0.3em @R=0.05em @! 
{
\lstick{\ket 0} & \gate{H} & \ctrl{2} & \ctrl{1} & \gate{H} & \meter & \qw 
\\
\lstick{\ket 0^{\otimes n}} & \gate{S(\boldsymbol{\theta}^k)} & \qw & \ctrl{1} & \qw & \qw & \qw 
\\
\lstick{\ket 0^{\otimes n}} & \qw & \gate{I^{k-1}} & \targ & \qw & \qw & \qw \\
}
\]
\caption{}
\end{subfigure}

 \vspace{0.5cm}

\begin{subfigure}[]{\textwidth}
\[
\Qcircuit @C=1.4em @R=0.6em
{
\lstick{\ket 0} & \gate{H} & \ctrlo{3} & \ctrl{3} & \gate{H} & \meter & \qw\\
& & & & & \\
& & & & & \mathcal{H} \\
\lstick{\ket 0^{\otimes n}} & \qw & \gate{U(\boldsymbol{\theta}^k)} & \gate{U^{k-1}} & \gate{\mathcal{I}/\mathcal{S}} & \meter & \qw
}
\]
\caption{}
\end{subfigure}

 \vspace{0.5cm}

\caption{ Circuits for measuring (a) overlap $\langle u(\boldsymbol{\theta}^k), u^j \rangle$ and (b) expectation $\bra{U(\boldsymbol{\theta}^k)}A\ket{U(\boldsymbol{\theta}^k)}$ (\cref{eq:cost_function}), (c) nonlinear overlap $\langle u^k,\Lambda_{\pm}^{k-1} u^{k-1}\rangle$ (\cref{eq:burger_cost}), (d) expectation  $\bra{S(\boldsymbol{\theta}^k)}\Lambda^{k-1}_I \ket{S(\boldsymbol{\theta}^k)}$  (\cref{eq:SEIR_CS}) and (e) higher-order overlap $\langle u^k,B u^{k-1}\rangle$ (\cref{eq:CN_cost}). $U(\boldsymbol{\theta}^k)$ denotes an ansatz with parameters $\boldsymbol{\theta}$ optimized at time $k$ and $U^j$ denotes an ansatz with parameters fixed at time $j$. $\mathcal{I}/\mathcal{S}$ denotes an optional shift operator (\cref{eq:shift_op}) and $\mathcal{S}/\mathcal{S}^{\dagger}$ denotes either incremental or decremental shift operator. Operators $S$ and $I$ are the ansaetze representing the respective variables.}
\label{fig:figA2}
\end{figure}
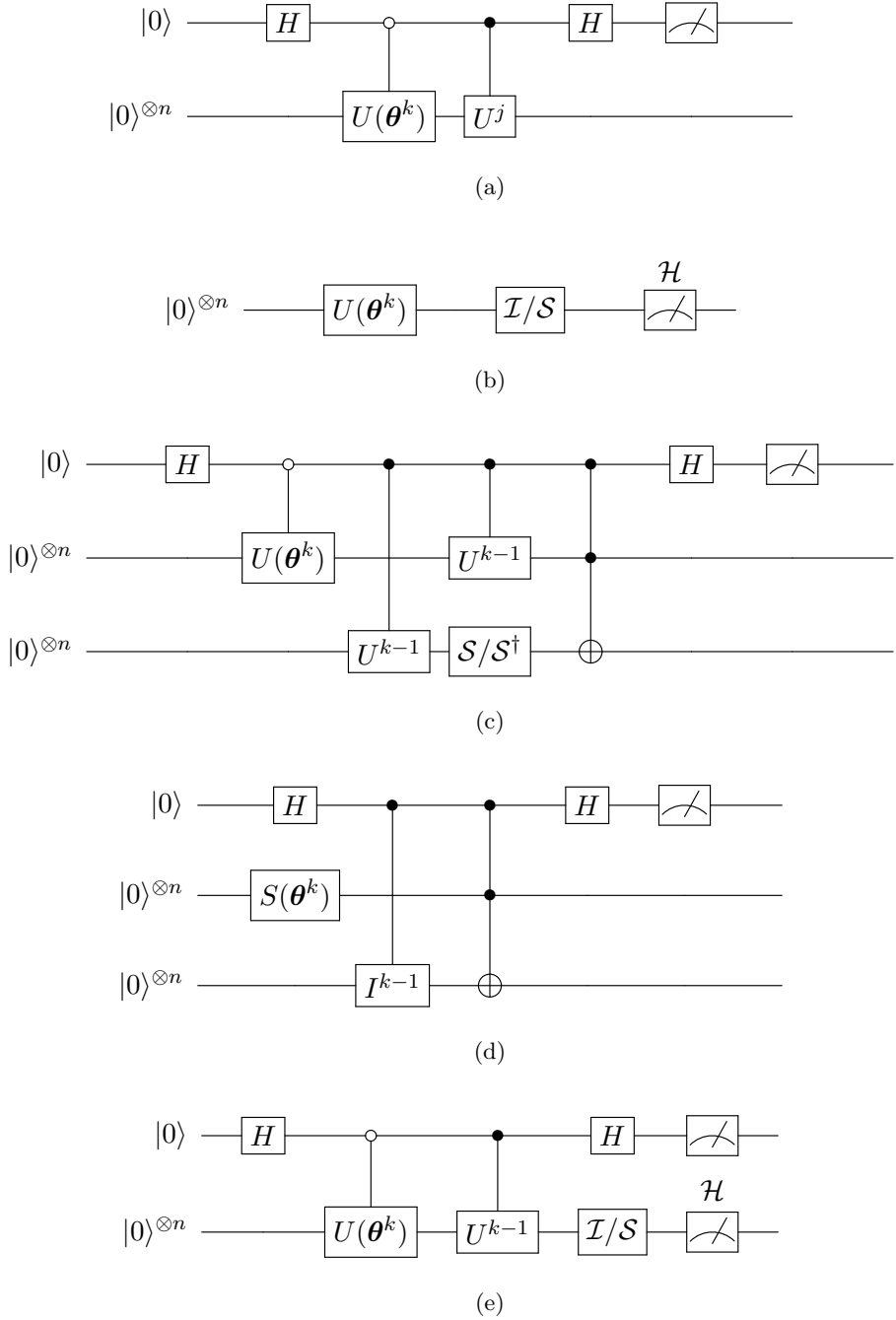

In this appendix, we give explicit circuit diagrams for some of the circuits we used in this paper. Real-amplitude hardware-efficient ansaetze are shown in \cref{fig:figA1}. Circuits used for measuring expectations and overlaps are shown in \cref{fig:figA2}.

\section{Higher-order Crank-Nicolson scheme} \label{sec:C} 
The quadratic scaling of quantum complexity in time steps Eq.~\eqref{eq:time_scaling} motivates higher-order time-stepping methods in order to limit time discretization and its errors. The Crank-Nicolson method is a second-order finite-difference scheme that retains the stability advantage associated with implicit methods \cite{Ali2017}. Using the trapezoidal rule, the finite difference of the spatial derivative Eq.~\eqref{eq:spatial_finite_difference} takes the midpoint between forward and backward Euler schemes, yielding
\begin{align}
    \partial_{xx} u(t_k,x_n) \simeq \frac{1}{2h^2}\left[ \left(u_{n+1}^k - 2u_{n}^k + u_{n-1}^k \right) + \left( u_{n+1}^{k-1} - 2u_{n}^{k-1} + u_{n-1}^{k-1} \right) \right].
\end{align}
Equating to the finite difference of the Caputo derivative, we obtain
\begin{align} 
    \left(1 - \frac{a}{2}\delta \right) u_{n}^k
    = \frac{a}{2}\delta u_n^{k-1} + w_k u_n^0  - \sum_{j=1}^{k-1} \Delta w_{j} u_n^{k-j},
\end{align}
or in matrix form,
\begin{align}
    A \mathbf{u}^k 
    = B \mathbf{u}^{k-1} + w_k \mathbf{u}^0 - \sum_{j=1}^{k-1} \Delta w_{j} \mathbf{u}^{k-j} ,
\end{align}
where $A = \mathbb{I} - B \in \mathbb{R}^{N \times N}$, $B = (a/2)\mathcal{L} \in \mathbb{R}^{N \times N}$, $\mathcal{L}$ is a symmetric tridiagonal matrix with $-2$ along its main diagonal and $1$ along the adjacent off-diagonals, and $\mathbb{I}$ is the identity matrix. This can be iterated using variational quantum optimization \cref{eq:fde} as before, with an updated cost function 
\begin{align}
    \mathcal{C}^k = -\frac{1}{2} \frac{\left( r^{k-1} \langle u^k,B u^{k-1} \rangle + \langle u^k, \tilde{f}^{k-1} \rangle \right)^2} 
    {\bra{u^k} A \ket{u^k}},
    \label{eq:CN_cost}
\end{align}
which includes a new term $\langle u^k,B u^{k-1}\rangle$ to be measured as a sum of observables ($2-4$, Eq.~\eqref{eq:A_decomposition}), using the quantum circuit (\cref{fig:figA2}e). 

\section{Truncated Caputo integral} \label{sec:D}
Since the coefficient $\Delta w_{j}$ of the $(k-j)$-th term decreases monotonically with $j$ for constant fractional power index $\alpha$, the $\mathcal{O}(M^2)$ time complexity scaling Eq.~\eqref{eq:time_scaling} can be limited to $\mathcal{O}(M)$ by setting an upper bound to the limits of the Caputo integral. This is done by replacing the Caputo finite difference derivative in Eq.~\eqref{eq:caputo_FD_derivative} with
\begin{align}
    D_t^\alpha u(t_k,x_n) \simeq g_{\alpha,\tau} \left[ w_1 u_n^k + \sum_{j=1}^{\min\{k,\xi\}} \left(w_{j+1}-w_{j}\right)u_n^{k-j} \right],
\end{align}
where $\xi \in [1,M]$ and $w_{k+1}=0$. For $\xi < k$, the truncation error thus incurred is
\begin{align}
    \varepsilon_{\tr}(\alpha, \xi) \sim \mathcal{O} \left[ g_{\alpha,\tau} (\Delta w_{\xi + 1}) r^{k-(\xi+1)} \right],
\end{align}
where $\Delta w_{\xi+1} = w_{\xi+2}-w_{\xi+1}$. A similar trade-off between solution fidelity and time complexity applies to classical finite difference schemes.

\end{appendices}

\section*{Acknowledgements}
We acknowledge the use of IBM Quantum services for this work. The views expressed are those of the authors, and do not reflect the official policy or position of IBM or the IBM Quantum team. This research is supported by the National Research Foundation, Singapore and the Agency for Science, Technology and Research (A*STAR) under the Quantum Engineering Programme (NRF2021-QEP2-02-P03), and A*STAR C230917003. This project is supported by the Singapore Ministry of Health’s National Medical Research Council through the Programme for Research in Epidemic Preparedness and Response (PREPARE), under Environmental Transmission \& Mitigation Co-operative (PREPARE-CS1-2022-004). DEK acknowledges funding support from the A*STAR Central Research Fund (CRF) Award for Use-Inspired Basic Research. DP acknowledges support from the Ministry of Education Singapore, under the grant MOE-T2EP50120-0019. 

\section*{Author Declarations}
The authors have no conflicts to disclose.

\section*{Data Availability}
The data that support the findings of this study are available from the corresponding author upon reasonable request. 

\bibliographystyle{unsrt}
\bibliography{bib}

\begin{thebibliography}{10}
\bibitem{comment2}
For example, in decreasing index order, we have $\prod_{i=3}^1 {A_i = A_3 A_2
  A_1}$.
\end{thebibliography}

\end{document}